\documentclass[onecolumn,showpacs,preprintnumbers,amsmath,amssymb]{revtex4}

\usepackage{graphicx}
\usepackage{dcolumn}
\usepackage{bm}
\usepackage{url}
\usepackage{color}
\usepackage{booktabs}
\usepackage[hidelinks]{hyperref}
\usepackage[nameinlink,noabbrev]{cleveref}

\begin{document}


\title{Inverse-mapped density-dependent relativistic mean-field inference of the neutron-star equation of state with multi-messenger constraints}



\author{Wen-Jie Xie}
\email{xiewenjie@ycu.edu.cn}
\affiliation{Department of Physics, Yuncheng University, Yuncheng 044000, China}
\affiliation{Shanxi Province Intelligent Optoelectronic Sensing Application Technology Innovation Center, Yuncheng University, Yuncheng 044000, China}
\author{Cheng-Jun Xia}
\email{cjxia@yzu.edu.cn}
\affiliation{Center for Gravitation and Cosmology, College of Physical Science and Technology, Yangzhou University, Yangzhou 225009, China}

\begin{abstract}
We present a Bayesian inference of the equation of state (EOS) of cold, dense matter within a density-dependent relativistic mean-field (DD-RMF) model. To connect macroscopic nuclear-matter properties to microscopic interaction channels in a reproducible way, we employ an explicit inverse-mapping procedure that constructs the density-dependent couplings from a physically interpretable 10-dimensional parameter set, while enforcing thermodynamic consistency and applying stability and causality filters. The reconstructed EOS is confronted with complementary multi-messenger constraints: low-density chiral effective field theory ($\chi$EFT) bands, intermediate-density heavy-ion collision (HIC) flow information, NICER mass--radius posteriors, and the existence of $\sim 2\,M_\odot$ pulsars. We find that the combined dataset strongly restricts both isovector and isoscalar sectors. In particular, the $\chi$EFT band favors a relatively soft symmetry-energy slope, $L\simeq 38~\mathrm{MeV}$, which correlates with a compact canonical radius $R_{1.4}\simeq 11.6~\mathrm{km}$. To simultaneously reproduce the intermediate-density softness indicated by HIC constraints and the high-density stiffness required by heavy pulsars, the posterior favors a moderately large Dirac effective mass at saturation ($M^*/M\simeq 0.64$) together with correlated, non-vanishing high-density limits of the scalar and vector couplings. The inferred sound-speed profile remains causal and exhibits pronounced nonconformal stiffening, with $c_s^2$ exceeding $1/3$ around a few times saturation density ($n\sim 3n_0$), suggesting that matter in the cores of massive neutron stars is far from a scale-invariant conformal regime. Finally, evidence-based diagnostics yield strong overall compatibility (e.g., $\ln R\gtrsim 9$ for the combined NS+$\chi$EFT+HIC sectors) within the present DD-RMF model class and adopted priors, indicating that terrestrial and astrophysical constraints can be jointly accommodated in a unified description of the neutron-star EOS.
\end{abstract}

\maketitle
\section{Introduction}
\label{sec:intro}
Determining the equation of state (EOS) of cold, neutron-rich matter from subnuclear densities to several times nuclear saturation density is a central problem at the interface of nuclear theory, terrestrial experiments, and neutron-star astrophysics. 
The EOS controls the structure of neutron stars through the Tolman--Oppenheimer--Volkoff (TOV) equations and imprints itself on a range of observables, including masses, radii, and tidal responses in compact-binary inspirals \cite{OppenheimerVolkoff1939,Hinderer2008,FlanaganHinderer2008}. 
Over the past decade, the growing set of multimessenger and electromagnetic constraints has enabled increasingly precise EOS inference. Simultaneously, this influx of data has highlighted the critical need for microphysically interpretable models capable of consistently connecting low-density nuclear constraints to the extreme high-density regime realized in neutron-star cores \cite{OzelFreire2016ARAA,LattimerPrakash2016PhysRep,Oertel2017RMP}.

On the astrophysical side, the existence of $\sim 2\,M_\odot$ neutron stars sets a robust lower bound on the stiffness of the EOS at high densities. 
The first precise measurement of a two-solar-mass pulsar was reported for PSR~J1614$-$2230, and subsequent observations have confirmed similarly massive objects, including PSR~J0740$+$6620 \cite{Demorest2010Nature,Cromartie2020NatAstron,Fonseca2021ApJL}. 
In parallel, pulse-profile modeling with NICER has provided direct mass--radius information for individual pulsars. 
Early NICER analyses of PSR~J0030$+$0451 established the feasibility of precise radius inference, followed by NICER$+$XMM constraints for the high-mass source PSR~J0740$+$6620 and subsequent reanalyses with improved background and expanded exposure \cite{Miller2019J0030,Riley2019J0030,Riley2021J0740,Miller2021J0740,Raaijmakers2021EOS,Salmi2022J0740,Salmi2024J0740}. 
More recently, NICER has delivered a radius constraint for the nearby millisecond pulsar PSR~J0437$-$4715, which has already been incorporated into updated EOS inference studies \cite{Choudhury2024J0437,BrandesWeise2025PRD,Koehn2025PRX}. 
Complementary constraints arise from gravitational-wave observations: the discovery event GW170817 and subsequent EOS-focused analyses provided landmark bounds on tidal deformabilities and radii, with additional information coming from later events such as GW190425 and the broader catalog of compact-binary coalescences \cite{Abbott2017GW170817,Abbott2018GW170817Radii,Abbott2020GW190425,Abbott2023GWTC3}. Furthermore, reconciling the presence of these massive pulsars with theoretical bounds from perturbative quantum chromodynamics (pQCD) has sparked intense debate regarding the maximum speed of sound in dense matter and whether it breaches the conformal limit ($c_s^2 = 1/3$) \cite{Bedaque2015, Annala2020, Altiparmak2022}.

Terrestrial nuclear-physics constraints probe fundamentally different density and isospin regimes, playing a crucial role in breaking degeneracies inherent to purely astrophysical inference. 
At densities up to $\sim (1\!-\!2)n_0$, chiral effective field theory ($\chi$EFT) provides a systematically improvable description of nuclear interactions and has become a cornerstone for low-density neutron-star matter constraints \cite{Hebeler2010PRL,Hebeler2013ApJ,DrischlerHoltWellenhofer2021ARNPS}. 
Recent efforts have emphasized uncertainty quantification and correlations in the nuclear-matter EOS, developing high-fidelity $\chi$EFT-based EOS constructions across proton fraction and temperature using statistical emulation techniques \cite{Drischler2020PRC,Keller2023PRL}. 
These developments enable more informative priors and tighter low-density bands for Bayesian EOS inference, directly informing recent analyses of the symmetry energy and neutron-star properties \cite{LimSchwenk2024PRC,Rutherford_2024}. 
At intermediate and higher densities, heavy-ion collisions (HIC) offer sensitivity to the pressure of nuclear matter over a broad range of thermodynamic conditions. 
Classic flow analyses constrained the symmetric-matter EOS, while more recent experiments have targeted the symmetry energy at suprasaturation density \cite{Danielewicz2002Science,Russotto2016ASYEOS}.  
A comprehensive synthesis of HIC-based EOS constraints and their connection to neutron-star physics has been provided in a recent community white paper, and Bayesian joint analyses have demonstrated the vital complementarity between HIC information and multimessenger data \cite{Sorensen2024PPNP,Huth2022Nature}. 
Additional nuclear structure information is supplied by parity-violating electron scattering measurements of neutron skins, including PREX-2 for $^{208}$Pb and CREX for $^{48}$Ca, which constrain isovector physics near saturation and further inform the density dependence of the symmetry energy \cite{Adhikari2021PREX2,Adhikari2022CREX}.

A central modeling challenge is therefore to construct an EOS framework that is mathematically flexible enough to accommodate these combined constraints across varied densities and isospins, while remaining thermodynamically consistent, strictly causal, and microphysically interpretable. 
Covariant density functional theory in the relativistic mean-field (RMF) approximation provides a natural platform in this respect. 
The foundational Walecka model and its nonlinear extensions established the basic mechanism of large scalar and vector self-energies and their interplay in nuclear saturation \cite{Walecka1974,BogutaBodmer1977}. 
Subsequent developments clarified the role of covariant mean fields in finite nuclei and nuclear matter, leading to widely used parametrizations (e.g., NL3 and FSUGold) that demonstrate the phenomenological success of calibrated RMF interactions \cite{Ring1996,SerotWalecka1997,Lalazissis1997NL3,ToddRutelPiekarewicz2005FSUGold}. 
In density-dependent RMF (DD-RMF) models, complex medium effects are captured through explicit density dependence of the effective couplings, while thermodynamic consistency is rigorously maintained via the rearrangement contribution. 
The Typel--Wolter form introduced a compact rational parametrization of these couplings, underlying several high-performance covariant energy density functionals \cite{TypelWolter1999,Lalazissis2005DDME2,Niksic2008DDPC1}. 
Nonetheless, extrapolating the isovector sector and the suprasaturation behavior of the couplings remains a dominant uncertainty for neutron-star applications. This motivates inference strategies capable of isolating exactly which density regions and interaction channels are being constrained by specific datasets.

In this work, we develop a comprehensive Bayesian inference framework based on a DD-RMF EOS. Crucially, the density dependence of the couplings is not sampled as an arbitrary unconstrained function. 
Instead, we employ an exact inverse-mapping strategy: a compact set of physically interpretable saturation properties (e.g., $K_0$, $M_0^*/M$, $E_{\rm sym}(n_0)$, $L$, $K_{\rm sym}$) together with minimal high-density asymptotic control parameters uniquely specifies the full density dependence of the isoscalar and isovector couplings in a Typel--Wolter--type rational form \cite{TypelWolter1999,Lalazissis2005DDME2}. 
This approach renders the inferred coupling functions directly interpretable in terms of fundamental nuclear-matter systematics and enables a transparent decoupling of low-density, intermediate-density, and high-density physics. 
We confront the resulting EOS with selectable combinations of neutron-star mass--radius information, gravitational-wave constraints, $\chi$EFT bands, and HIC-based pressure constraints, while strictly enforcing thermodynamic consistency and causality \cite{Miller2019J0030,Riley2019J0030,Riley2021J0740,Miller2021J0740,Salmi2024J0740,Choudhury2024J0437,Abbott2017GW170817,Abbott2018GW170817Radii,Abbott2023GWTC3,Hebeler2010PRL,DrischlerHoltWellenhofer2021ARNPS,Keller2023PRL,Danielewicz2002Science,Sorensen2024PPNP}. 
Posterior exploration is performed with nested sampling using MultiNest \cite{Skilling2004NestedSampling,Feroz2009MultiNest}. 
By evaluating the Bayesian evidence for each data combination, this framework enables rigorous diagnostics of complementarity and tension among the diverse constraints. Ultimately, it yields a quantified reconstruction of density-dependent RMF couplings that seamlessly unifies current nuclear and astrophysical information into a coherent picture of cold, dense matter \cite{Huth2022Nature,Koehn2025PRX,BrandesWeise2025PRD,Malik2022_ApJ930-17,Traversi2020PRC}.

The remainder of this paper is organized as follows. In Sec.~\ref{sec:theory} we detail the DD-RMF model and the inverse mapping that reconstructs density-dependent couplings from the chosen parameter set, describe the EOS construction under $\beta$ equilibrium, and summarize the likelihood components and Bayesian evidence diagnostics. 
Section~\ref{sec:results} presents posterior constraints from individual and combined data sets, reconstructed coupling functions and EOS bands, and an evidence-based assessment of dataset compatibility. 
We conclude in Sec.~\ref{sec:summary} with a summary and outlook.

\section{Theoretical framework}
\label{sec:theory}

\subsection{Density-dependent relativistic mean-field model and inverse mapping}
\label{sec:theory_rmf}

We adopt a density-dependent relativistic mean-field (DD-RMF) energy-density functional for cold, uniform matter. 
Relativistic mean-field models provide a compact and microphysically interpretable description of nuclear saturation in terms of large scalar and vector self-energies \cite{Walecka1974,BogutaBodmer1977,SerotWalecka1997,Ring1996}. 
In DD-RMF, medium effects are encoded through explicit density dependence of the effective couplings while preserving thermodynamic consistency via the rearrangement contribution \cite{TypelWolter1999,Typel2005PRC,Xia2024_PRD110-114009}. 
For broad context on covariant parameterizations under empirical nuclear-matter constraints, see Ref.~\cite{Dutra2014RMFConstraints}; throughout this work, we use natural units $\hbar=c=1$.

For uniform matter composed of neutrons ($n$) and protons ($p$) with number densities $n_n$ and $n_p$, we define the total baryon density $n \equiv n_n+n_p$, the isospin asymmetry $n_3 \equiv n_p-n_n$, and the proton fraction $x_p \equiv n_p/n$. 
The Dirac effective mass is expressed as $M^*(n_n,n_p) = M - \Sigma_S$, where $\Sigma_S = G_\sigma(n)\,n_s$ is the scalar self-energy. 
The total scalar density is given by $n_s = \sum_{i=n,p} n_{s,i}$, with
\begin{equation}
n_{s,i}=\frac{2}{(2\pi)^3}\int_0^{k_{F,i}}\!d^3k\,\frac{M^*}{\sqrt{k^2+M^{*2}}}.
\label{eq:scalar_density}
\end{equation}
The kinetic (nucleonic) energy density $\varepsilon_N$ and pressure $P_N$ are evaluated via the standard integrals over the Fermi sea:
\begin{align}
\varepsilon_N &= \sum_{i=n,p}\frac{2}{(2\pi)^3}\int_0^{k_{F,i}}\!d^3k\,\sqrt{k^2+M^{*2}},
\label{eq:ekin}\\
P_N &= \sum_{i=n,p}\frac{2}{3(2\pi)^3}\int_0^{k_{F,i}}\!d^3k\,\frac{k^2}{\sqrt{k^2+M^{*2}}}.
\label{eq:pkin}
\end{align}
The interaction (mean-field) contribution is written in a compact point-coupling form,
\begin{equation}
\varepsilon_{\rm int}=
\frac12\,G_\sigma(n)\,n_s^2+\frac12\,G_\omega(n)\,n^2+\frac12\,G_\rho(n)\,n_3^2,
\label{eq:eps_int}
\end{equation}
which is equivalent to the usual meson-exchange formulation at the mean-field level after identifying the effective couplings $G_i(n)\equiv g_i^2(n)/m_i^2$ for the $\sigma$, $\omega$, and $\rho$ mesons \cite{TypelWolter1999,Lalazissis2005DDME2,Niksic2008DDPC1}. 
Crucially, the density-dependent couplings introduce a rearrangement self-energy \cite{TypelWolter1999,Typel2005PRC},
\begin{equation}
\Sigma_R(n)=
\frac12\,n^2\frac{dG_\omega}{dn}
-\frac12\,n_s^2\frac{dG_\sigma}{dn}
+\frac12\,n_3^2\frac{dG_\rho}{dn},
\label{eq:sigmaR}
\end{equation}
which must be included in both the pressure and the chemical potentials to maintain thermodynamic consistency. 

To describe the density dependence of the couplings, we write $G_i(n)=G_{i0}\,f_i^2(n)$ for $i\in\{\sigma,\omega,\rho\}$, where $G_{i0}\equiv G_i(n_0)$ is the coupling strength at saturation density, and the shape functions $f_i(n)$ satisfy $f_i(n_0)=1$. 
Following the Typel--Wolter ansatz \cite{TypelWolter1999}, we adopt a rational form:
\begin{equation}
f_i(n) = a_i\,\frac{1+b_i\left(x+d_i\right)^2}{1+c_i\left(x+d_i\right)^2},
\qquad x\equiv n/n_0,
\label{eq:tw_form}
\end{equation}
which naturally approaches an asymptotic limit $f_{i,\infty}\equiv \lim_{x\to\infty} f_i(x)=a_i(b_i/c_i)$ at high densities.

Instead of freely varying the microscopic coefficients $a_i, b_i, c_i$, and $d_i$, our inference is performed in a physically intuitive 10-dimensional macroscopic parameter space:
\begin{equation}
\bm{\theta}=
\left(K_0,\; (M^*/M)_0,\; n_0,\; E_0,\; E_{\rm sym,0},\; L,\; K_{\rm sym},\;
f_{\sigma,\infty},\; f_{\omega,\infty},\; f_{\rho,\infty}\right).
\label{eq:theta_def}
\end{equation}
Here, $E_0\equiv (E/A)(n_0)-M$ is the binding energy per nucleon at saturation. 
These empirical parameters are defined in the standard way by expanding the energy per baryon $E(n,\delta)$ around saturation with respect to density and the isospin asymmetry parameter $\delta\equiv (n_n-n_p)/n$ \cite{LiChenKo2008IsospinPR,Tsang2012SymConstraints}.

To connect this macroscopic parameter space to the microscopic effective couplings, we employ an exact inverse mapping procedure. 
The coupling strengths at saturation ($G_{\sigma0}, G_{\omega0}, G_{\rho0}$) are analytically fixed by the effective mass, binding energy, and symmetry energy, respectively. 
The density-dependence shape parameters are subsequently determined by matching the specified isovector properties ($L, K_{\rm sym}$) and isoscalar properties ($K_0$ and the zero-pressure condition at saturation). 
The detailed step-by-step mathematical derivation of this inverse mapping is provided in Appendix \ref{app:inverse_mapping}. 
This rigorous mapping guarantees that every sampled EOS perfectly matches the input nuclear matter properties.

\subsection{Equation of state and neutron-star structure}
\label{sec:theory_eos}

With the model parameters determined, we compute the core EOS by imposing cold $\beta$ equilibrium and charge neutrality at each baryon density $n$: $\mu_n=\mu_p+\mu_e$, $\mu_e=\mu_\mu$, and $n_p=n_e+n_\mu$. 
The nucleon chemical potentials explicitly include the rearrangement term:
\begin{equation}
\mu_{p,n} = E_{F,n,p}^* + G_\omega(n)\,n \mp G_\rho(n)\,n_3 + \Sigma_R(n),
\label{eq:chem_pot}
\end{equation}
where $E_{F,i}^*=\sqrt{k_{F,i}^2+M^{*2}}$, and the leptons ($e, \mu$) are treated as free Fermi gases. 
For a given density $n$, we solve for the proton fraction $x_p(n)$ and subsequently evaluate the total energy density $\varepsilon(n) = \varepsilon_N+\varepsilon_{\rm int}+\varepsilon_\ell$ and the total pressure $P(n)$. 
Numerically, we obtain the pressure directly from the thermodynamic relation $P(n)= n(d\varepsilon/dn)-\varepsilon(n)$ using finite-difference derivatives on the computed $\varepsilon(n)$ to guarantee exact thermodynamic consistency.

To describe the outer layers of the neutron star, we attach a tabulated crust EOS based on the SLy4 functional \cite{DouchinHaensel2001SLy4}. 
The transition point between the crust and the core is identified by enforcing continuity and monotonicity of $P(\varepsilon)$ across the interface \cite{LattimerPrakash2016PhysRep,Oertel2017RMP}. 
Furthermore, we rigorously filter unphysical EOS models by enforcing causality, requiring that the squared speed of sound $c_s^2=dP/d\varepsilon \le 1$ throughout the entire density range.

The macroscopic structure of a non-rotating neutron star is obtained by integrating the Tolman--Oppenheimer--Volkoff (TOV) equations \cite{Tolman1939,OppenheimerVolkoff1939}:
\begin{align}
\frac{dP}{dr} &= -\frac{(\varepsilon+P)\left[m(r)+4\pi r^3 P\right]}{r\left[r-2m(r)\right]},
\label{eq:tov_P}\\
\frac{dm}{dr} &= 4\pi r^2 \varepsilon(r),
\label{eq:tov_m}
\end{align}
subject to the central boundary conditions $m(0)=0$ and $P(0)=P_c$. 
The stellar radius $R$ is defined at the surface where $P(R)=0$, yielding the gravitational mass $M=m(R)$. 
Scanning over a grid of central pressures produces the stable $M$--$R$ sequence and identifies the maximum mass $M_{\max}$ \cite{OzelFreire2016ARAA,LattimerPrakash2016PhysRep}. 

\subsection{Bayesian inference and observational constraints}
\label{sec:theory_bayes}

We employ Bayesian inference to confront the generated EOS models with multi-messenger data\cite{Xie2019_ApJ883-174,Xia2024_PRD110-114009,Xie2024PRD110_043025,xie2026prd}. 
The posterior distribution $p(\bm{\theta}|\mathcal{D})$ is proportional to the product of the prior $\pi(\bm{\theta})$ and the combined likelihood $\mathcal{L}(\bm{\theta})$, normalized by the Bayesian evidence $\mathcal{Z}$. 
We evaluate the posterior and evidence using the nested sampling algorithm \cite{Skilling2004NestedSampling,Skilling2006NestedSampling} implemented via \texttt{MultiNest} \cite{Feroz2009MultiNest}. 
The total log-likelihood is constructed as a modular sum of contributions from distinct data sectors:
\begin{equation}
\ln\mathcal{L} = \ln\mathcal{L}_{\rm NS} + \ln\mathcal{L}_{\chi{\rm EFT}} + \ln\mathcal{L}_{\rm HIC} + \ln\mathcal{L}_{M_{\max}}.
\label{eq:loglike_sum}
\end{equation}

For the astrophysical observations ($\ln\mathcal{L}_{\rm NS}$), we incorporate NICER pulse-profile constraints for PSR~J0030$+$0451 \cite{Miller2019J0030,Riley2019J0030}, PSR~J0740$+$6620 \cite{Riley2021J0740,Miller2021J0740,Salmi2024J0740}, and PSR~J0437$-$4715 \cite{Choudhury2024J0437}. 
For each source, the observational posterior in the $(R,M)$ plane is represented using a kernel density estimate (KDE), and the model agreement is quantified by integrating the KDE density along the predicted $M(R)$ curve. 
Additionally, the robust existence of massive pulsars dictates the maximum-mass consistency ($\ln\mathcal{L}_{M_{\max}}$) \cite{Demorest2010Nature,Antoniadis2013J0348,Cromartie2020NatAstron,Fonseca2021ApJL}. 
Rather than applying a hard cutoff, we softly penalize EOSs whose $M_{\max}$ falls below a target heavy-pulsar mass by appending a Gaussian tail penalty, maintaining statistical rigor.

\begin{figure}[htb]
\centering
\includegraphics[width=0.9\linewidth]{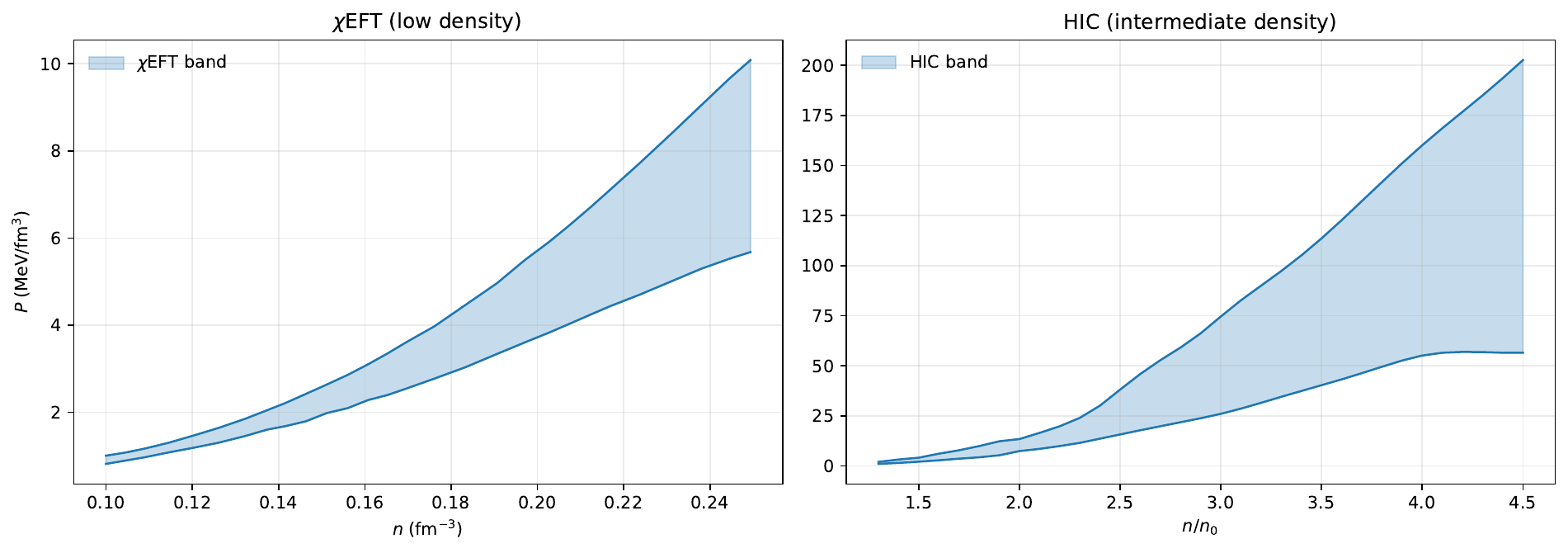} 
\caption{Microphysical constraints employed in the likelihood function. (Left) The chiral effective field theory ($\chi$EFT) pressure band evaluated for pure neutron matter at sub-saturation densities taken from Ref. \cite{Keller2023PRL}. (Right) The heavy-ion collision (HIC) pressure band constrained for symmetric nuclear matter at intermediate suprasaturation densities taken from Ref. \cite{Danielewicz2002_Science298-1592}.}
\label{fig:cheft_hic_bands}
\end{figure}

The microphysical constraints, visually summarized in Fig.~\ref{fig:cheft_hic_bands}, strongly guide the low- and intermediate-density behaviors of the EOS. 
At densities up to $\sim(1$--$2)n_0$, the predicted pressure $P(n)$ in neutron-rich matter is compared against a chiral EFT band ($\ln\mathcal{L}_{\chi{\rm EFT}}$), where deviations from the band midpoint are penalized with a Gaussian width defined by the band's theoretical uncertainty \cite{Hebeler2010PRL,Hebeler2013ApJ,DrischlerHoltWellenhofer2021ARNPS,Drischler2020PRC,Keller2023PRL,LimSchwenk2024PRC}. 
At intermediate densities, the symmetric-matter pressure $P_{\rm SNM}(n)$ is constrained by a heavy-ion collision band ($\ln\mathcal{L}_{\rm HIC}$), incorporating both classic flow bounds and modern isospin-sensitive syntheses \cite{Danielewicz2002Science,Russotto2016ASYEOS,Sorensen2024PPNP,Huth2022Nature}.

Finally, the nested sampling framework intrinsically computes the Bayesian evidence $\mathcal{Z}$ for each data configuration. 
Rather than performing traditional model selection, we utilize these evidence values to rigorously diagnose complementarity and tension among the diverse constraint sets within our model architecture. 
Specifically, we evaluate the conditional evidence to quantify the incremental information gain and Occam penalty when combining datasets. 
Furthermore, we compute the compatibility factor $\ln R$ \cite{KassRaftery1995,Trotta2008BayesSky}, which elegantly compares the hypothesis that two distinct datasets share the same underlying physical parameters against the hypothesis that they are statistically independent.

\section{Results and Discussions}
\label{sec:results}

\subsection{Posterior constraints on nuclear parameters}
\label{sec:results_parameters}

\begin{figure*}[t]
  \centering
  \includegraphics[width=\textwidth]{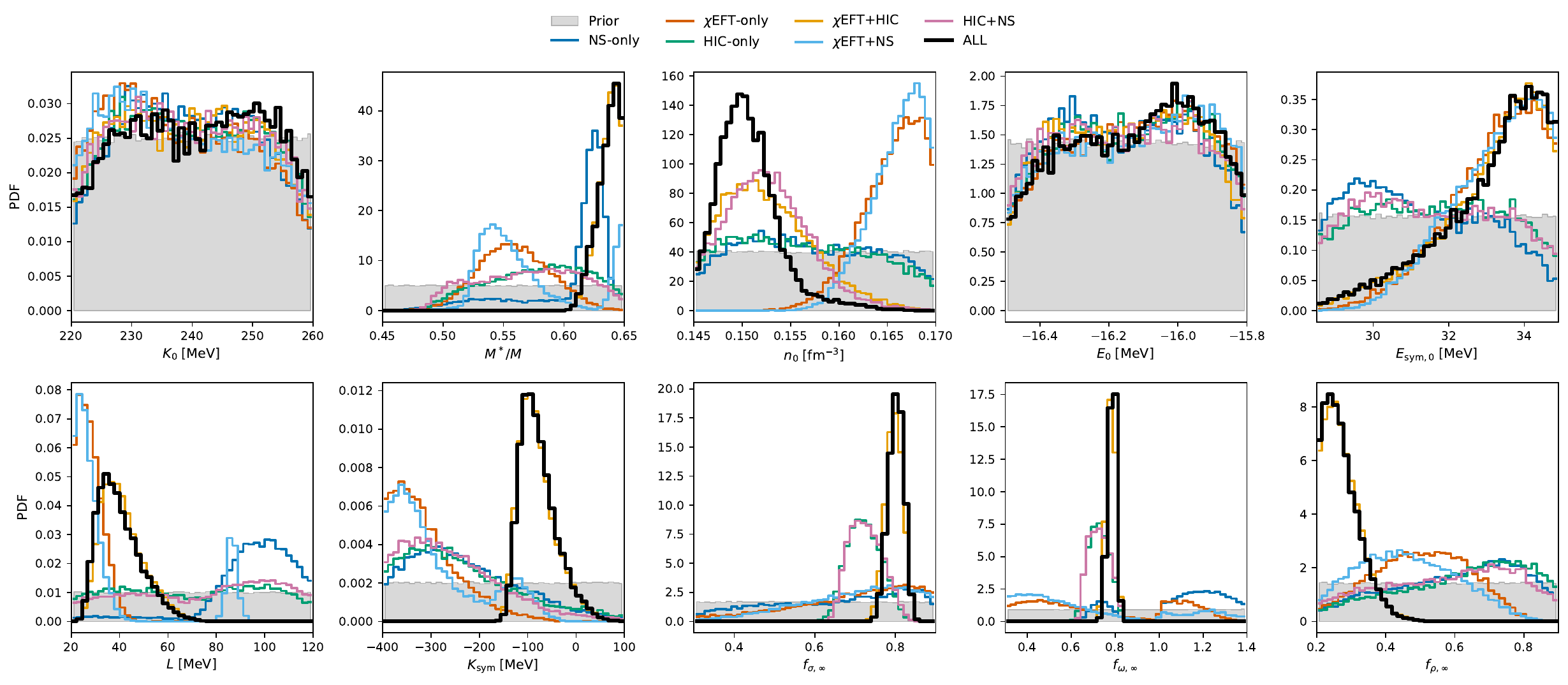}
  \caption{One-dimensional marginalized posteriors for the 10 inferred parameters under different data selections. The shaded gray band shows the uniform priors. Colored curves show posteriors obtained with NS-only, $\chi$EFT-only, HIC-only, $\chi$EFT+HIC, $\chi$EFT+NS, HIC+NS, and the full \texttt{ALL} combination.}
  \label{fig:1d_posteriors}
\end{figure*}

\begin{figure*}[t]
  \centering
  \includegraphics[width=\textwidth]{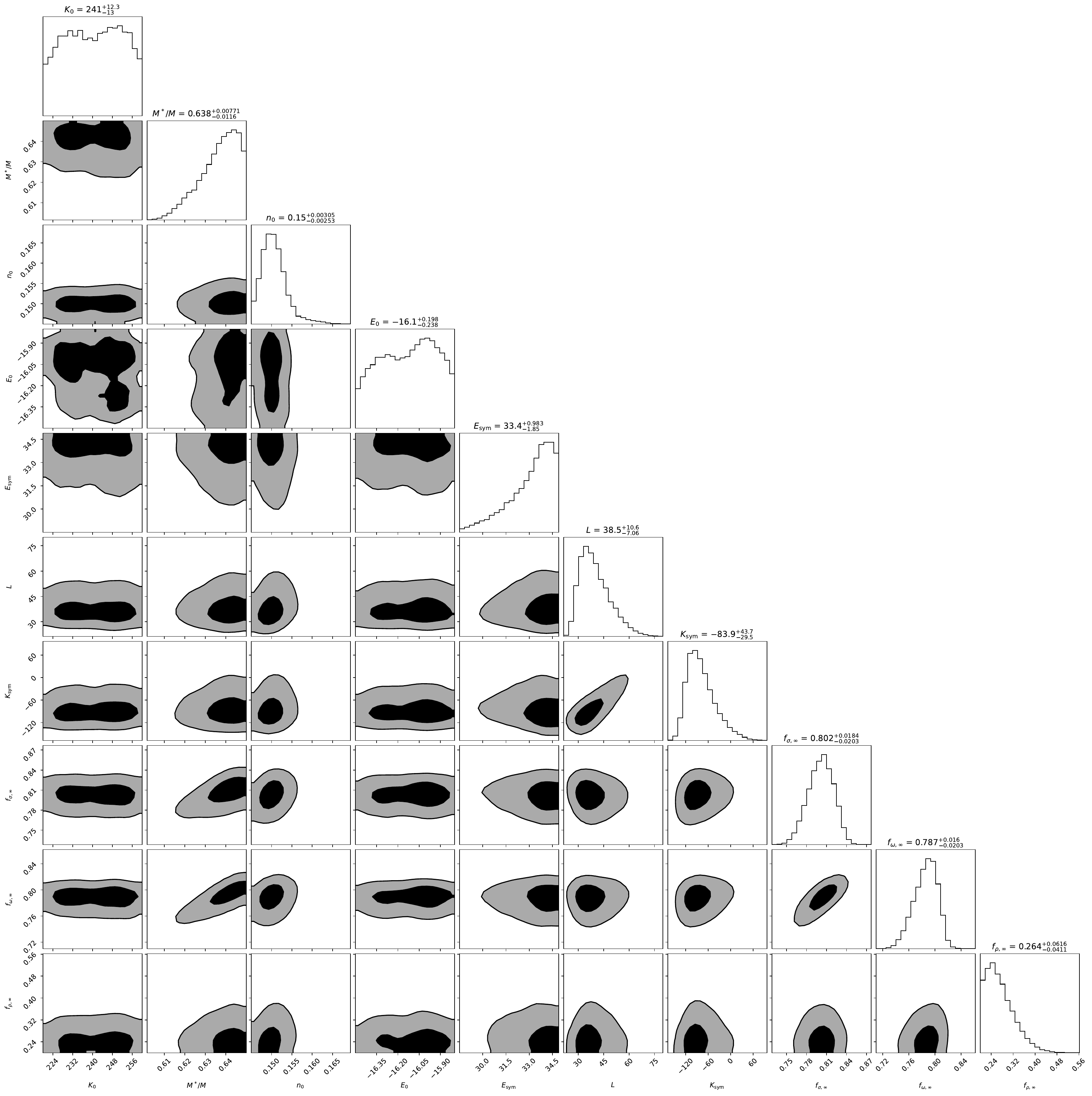}
  \caption{Corner plot for the \texttt{ALL} dataset. Diagonal panels show 1D marginalized posteriors, while off-diagonal panels show 2D joint posteriors. Inner (outer) shaded regions correspond to 68\% (90\%) credible regions.}
  \label{fig:corner_all}
\end{figure*}

We begin by examining the marginalized posterior distributions in the 10-dimensional nuclear-parameter space. Table~\ref{tab:alldata_ci} summarizes the uniform prior ranges and the 68\% and 90\% credible intervals (C.I.) for our reference combined analysis (hereafter \texttt{ALL}, comprising NS, $\chi$EFT, and HIC constraints together with the $M_{\max}$ penalty). To disentangle the information content of different constraint sectors, Fig.~\ref{fig:1d_posteriors} shows 1D marginalized posteriors for individual and partially combined datasets, while Fig.~\ref{fig:corner_all} displays the corresponding corner plot for \texttt{ALL}, highlighting correlations and degeneracies.

In the isoscalar sector, the incompressibility $K_0$ and saturation binding energy $E_0$ remain largely prior-dominated within our adopted ranges. This is not unexpected: the priors (e.g., $K_0\in[220,260]$~MeV) already encode substantial empirical guidance from nuclear experiments and theory \cite{Shlomo2006_EPJA30-23,Li2023PRL_ISGMR,Zhou2023PRC_NSInference,Xu2021PRC_K0Bayes}. The fact that the posteriors closely track these priors indicates that the present NS $M$--$R$ information and the HIC pressure bands, as implemented here, provide limited additional leverage on the saturation-point curvature $K_0$ or on $E_0$. In particular, HIC information enters mainly through pressure constraints at intermediate to suprasaturation densities, which does not necessarily translate into a tight determination of the curvature at saturation within our mapping.

In contrast, the Dirac effective mass $M^*/M$ and saturation density $n_0$ exhibit nontrivial competition among datasets. As shown in Fig.~\ref{fig:1d_posteriors}, $\chi$EFT-only and HIC-only inferences favor lower effective masses, $M^*/M\sim 0.55$, comparable to values typical of finite-nucleus-optimized interactions (e.g., DD-ME2 \cite{Lalazissis2005DDME2}). Meanwhile, astrophysical constraints---in particular the requirement that viable EOSs support heavy neutron stars (implemented via the $M_{\max}$ penalty)---favor parameter combinations that produce sufficient high-density stiffness. Within our DD-RMF inverse-mapping framework this preference correlates with a larger saturation effective mass, and the combined posterior peaks at $M^*/M\simeq 0.638$. While suprasaturation stiffness is ultimately controlled by the coupled evolution of scalar attraction and vector repulsion, $M^*/M$ serves here as a convenient saturation-scale indicator that is strongly correlated with the high-density behavior of the mean fields. Similarly, although NS information mildly prefers larger $n_0$ values, the $\chi$EFT band anchors the combined posterior near $n_0\simeq 0.150~\mathrm{fm}^{-3}$.

The isovector sector provides a clear demonstration of multi-messenger complementarity. The symmetry-energy slope $L$ and curvature $K_{\rm sym}$ are notoriously difficult to constrain from macroscopic NS $M$--$R$ data alone \cite{Oertel2017_RMP89-015007,Margueron2018PRC_MetaEOS_NS}. Consistent with this expectation, the NS-only posterior for $L$ in Fig.~\ref{fig:1d_posteriors} is broad. Once the microscopic $\chi$EFT band is included, this degeneracy is substantially reduced, yielding $L = 38^{+11}_{-7}$~MeV and $K_{\rm sym} = -84^{+44}_{-30}$~MeV at 68\% confidence level. The resulting soft isovector density dependence is consistent with recent \emph{ab initio} $\chi$EFT calculations \cite{Drischler2020PRC,Keller2023PRL} and aligns more naturally with the smaller neutron skin suggested by CREX on $^{48}$Ca \cite{Adhikari2022CREX}, while remaining in mild tension with the larger skin implied by PREX-II on $^{208}$Pb \cite{Adhikari2021PREX2}. The corner plot displayed in Fig.~\ref{fig:corner_all} further shows positive correlation between $L$ and $K_{\rm sym}$, reflecting familiar degeneracies of the symmetry energy around saturation preserved by our inverse-mapping parameterization.

\begin{table}[t]
\caption{Uniform prior ranges and posterior constraints for the 10 inferred parameters from the reference combined analysis (\texttt{ALL}: NS+$\chi$EFT+HIC+$M_{\max}$ penalty).}
\label{tab:alldata_ci}
\begin{ruledtabular}
\begin{tabular}{lccc}
Parameter & Prior range & 68\% C.I. & 90\% C.I. \\
\hline
  $K_0~(\mathrm{MeV})$ & $[220,\,260]$ & $241^{+12}_{-13}$ & $241^{+17}_{-18}$ \\
  $M^*/M$ & $[0.45,\,0.65]$ & $0.638^{+0.008}_{-0.012}$ & $0.638^{+0.010}_{-0.019}$ \\
  $n_0~(\mathrm{fm}^{-3})$ & $[0.145,\,0.170]$ & $0.1503^{+0.0031}_{-0.0025}$ & $0.1503^{+0.0061}_{-0.0039}$ \\
  $E_0~(\mathrm{MeV})$ & $[-16.5,\,-15.8]$ & $-16.12^{+0.20}_{-0.24}$ & $-16.12^{+0.27}_{-0.33}$ \\
  $E_{\rm sym,0}~(\mathrm{MeV})$ & $[28.5,\,34.9]$ & $33.43^{+0.98}_{-1.85}$ & $33.43^{+1.31}_{-3.23}$ \\
  $L~(\mathrm{MeV})$ & $[20,\,120]$ & $38^{+11}_{-7}$ & $38^{+19}_{-10}$ \\
  $K_{\rm sym}~(\mathrm{MeV})$ & $[-400,\,100]$ & $-84^{+44}_{-30}$ & $-84^{+80}_{-43}$ \\
  $f_{\sigma,\infty}$ & $[0.3,\,0.9]$ & $0.802^{+0.018}_{-0.020}$ & $0.802^{+0.029}_{-0.033}$ \\
  $f_{\omega,\infty}$ & $[0.3,\,1.4]$ & $0.787^{+0.016}_{-0.020}$ & $0.787^{+0.025}_{-0.034}$ \\
  $f_{\rho,\infty}$ & $[0.2,\,0.9]$ & $0.264^{+0.062}_{-0.041}$ & $0.264^{+0.110}_{-0.056}$ \\
\end{tabular}
\end{ruledtabular}
\end{table}

A particularly informative outcome of the inference concerns the high-density asymptotic control parameters $f_{\sigma,\infty}$, $f_{\omega,\infty}$, and $f_{\rho,\infty}$, which regulate suprasaturation behavior. Figure~\ref{fig:1d_posteriors} shows that neither NS data nor HIC data alone tightly constrains the isoscalar asymptotic limits ($f_{\sigma,\infty}$ and $f_{\omega,\infty}$), whereas their combination yields comparatively narrow posteriors,
$f_{\sigma,\infty} = 0.802^{+0.018}_{-0.020}$ and
$f_{\omega,\infty} = 0.787^{+0.016}_{-0.020}$ at 68\% confidence level.
The physical origin is visible in Fig.~\ref{fig:corner_all}, which exhibits a tight, approximately linear correlation between $f_{\sigma,\infty}$ and $f_{\omega,\infty}$. Satisfying intermediate-density HIC pressures, avoiding acausal stiffness (excluded by our physicality filters), and maintaining sufficient high-density support for $\sim 2\,M_\odot$ stars requires scalar attraction and vector repulsion to co-vary in a balanced manner. In addition, the posterior favors a suppressed isovector asymptotic limit, $f_{\rho,\infty} = 0.264^{+0.062}_{-0.041}$, indicating reduced $\rho$-channel repulsion in the deep core. This trend disfavors an excessively large high-density symmetry energy, which would otherwise increase proton fractions and could trigger early direct-Urca cooling in lower-mass stars, depending on the detailed composition and pairing \cite{LattimerPrakash2016PhysRep}.

\subsection{Reconstructed microphysics: density-dependent couplings}
\label{sec:results_couplings}

\begin{figure*}[t]
  \centering
  \includegraphics[width=\textwidth]{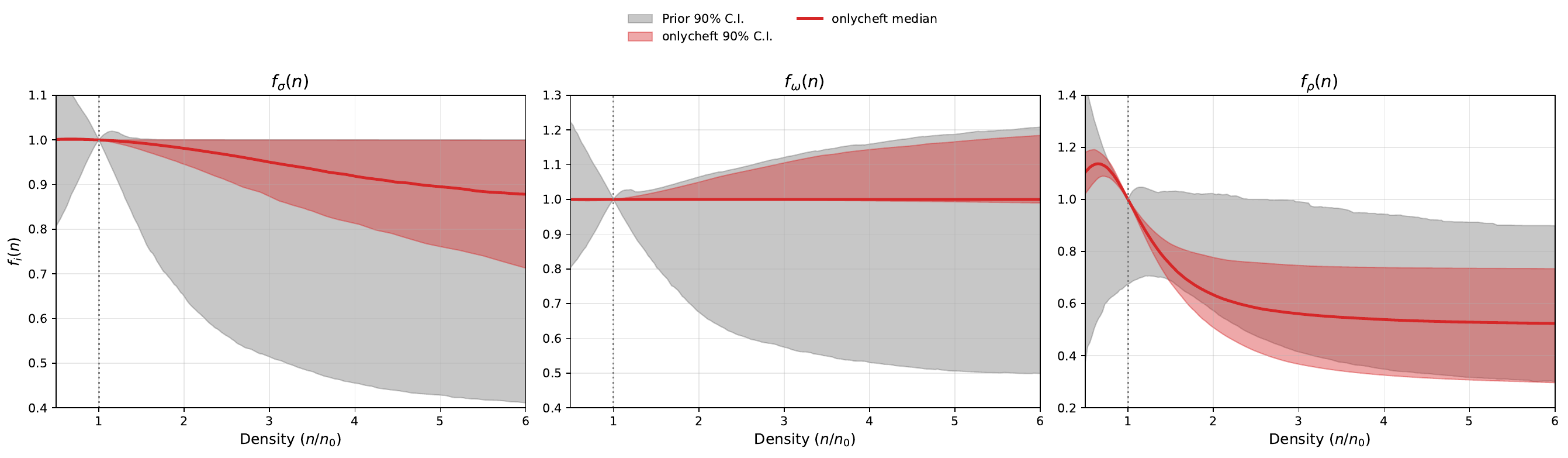}\\[1.0ex]
  \includegraphics[width=\textwidth]{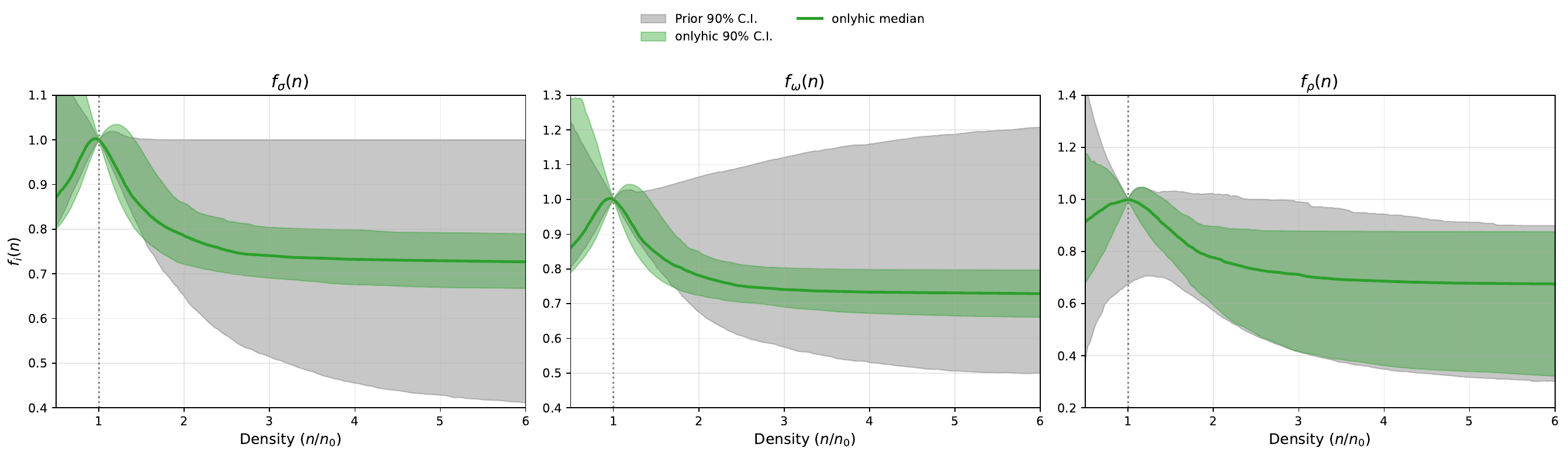}\\[1.0ex]
  \includegraphics[width=\textwidth]{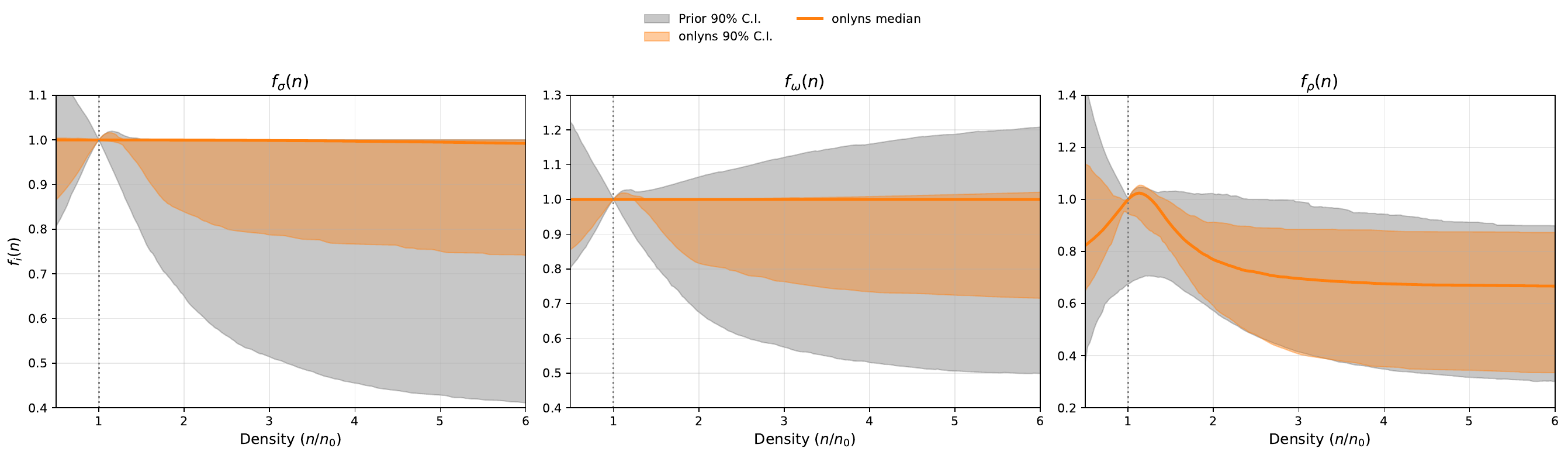}
  \caption{Reconstructed coupling shape functions $f_i(n)=g_i(n)/g_i(n_0)$ inferred from individual datasets (top: $\chi$EFT-only; middle: HIC-only; bottom: NS-only). In each row, the gray band denotes the 90\% prior interval and the colored band denotes the 90\% posterior interval. The vertical dotted line marks saturation density ($n=n_0$).}
  \label{fig:couplings_partial}
\end{figure*}

\begin{figure*}[t]
  \centering
  \includegraphics[width=\textwidth]{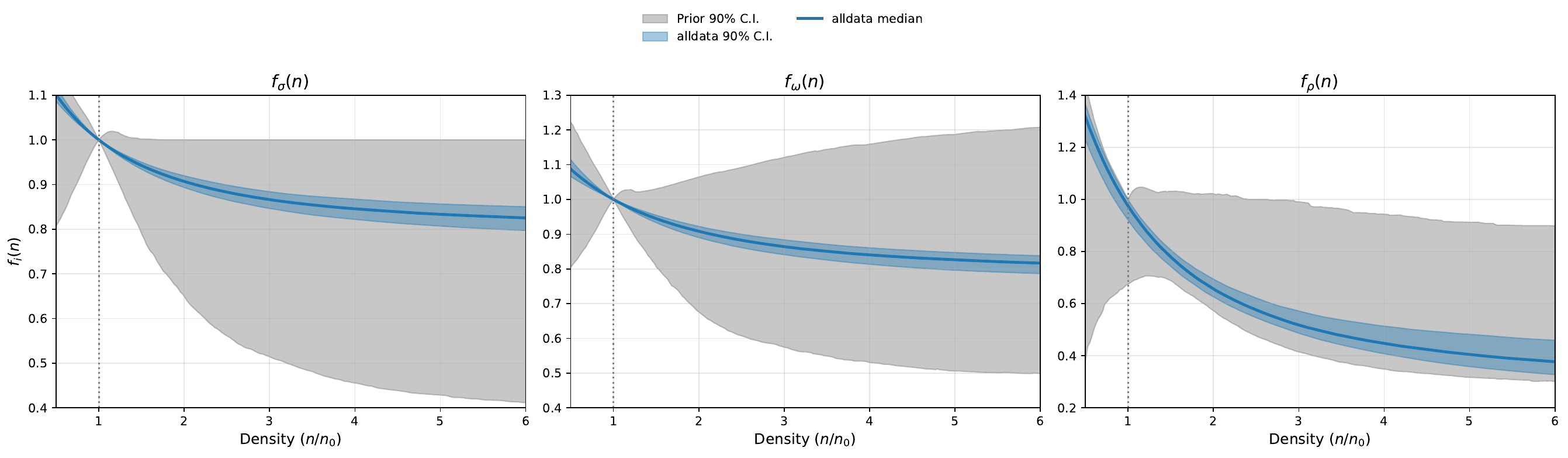}
  \caption{Reconstructed coupling shape functions inferred from the \texttt{ALL} dataset. The gray band shows the 90\% prior interval and the blue band shows the 90\% posterior interval. The multi-messenger combination yields narrow trajectories and exposes correlated isoscalar evolution together with a suppressed isovector channel at high density.}
  \label{fig:couplings_all}
\end{figure*}

We now map the inferred macroscopic parameters back to microscopic interaction channels by reconstructing the density-dependent coupling shape functions $f_i(n)=g_i(n)/g_i(n_0)$ for the scalar ($\sigma$), vector ($\omega$), and isovector ($\rho$) mesons. This reconstruction allows a direct diagnosis of which constraints control which channel over which density regime. Figure~\ref{fig:couplings_partial} shows the reconstructed 90\% posterior bands for individual datasets, while Fig.~\ref{fig:couplings_all} presents the \texttt{ALL} result.

The dataset-conditioned panels in Fig.~\ref{fig:couplings_partial} illustrate the distinct sensitivities and limitations of each probe. NS-only constraints largely leave the isoscalar coupling functions close to their prior ranges, reflecting a fundamental degeneracy: NS observables primarily constrain the bulk relation $P(\varepsilon)$ of $\beta$-equilibrated matter but do not uniquely determine its decomposition into scalar attraction and vector repulsion. $\chi$EFT-only constraints tighten the couplings at sub- and near-saturation densities ($n\lesssim 1.5\,n_0$) but rapidly lose constraining power at higher density where $\chi$EFT uncertainties grow. HIC-only constraints, derived mainly from SNM collective flow, preferentially restrict the isoscalar sector over intermediate densities ($\sim 2$--$4\,n_0$), while leaving the isovector channel relatively unconstrained.

The \texttt{ALL} inference represented in Fig.~\ref{fig:couplings_all} collapses these bands into narrow trajectories by jointly satisfying low-density $\chi$EFT microphysics, intermediate-density HIC pressures, and high-density NS structural demands. In the isoscalar sector, both $f_\sigma(n)$ and $f_\omega(n)$ show a gradual, correlated decline at suprasaturation density. This behavior is not imposed by the priors but emerges from the combined likelihood. The trend is consistent with the qualitative behavior of widely used density-dependent interactions such as DD-ME2 \cite{Lalazissis2005DDME2} and DD2 \cite{Typel2010_PRC81-015803}. The tight correlation between $f_\sigma$ and $f_\omega$ ensures that their large mean-field contributions partially cancel, leaving a net pressure that remains compatible with both HIC information and $M_{\max}\gtrsim 2\,M_\odot$, while maintaining thermodynamic consistency through the rearrangement term.

In the isovector channel, the reconstructed $f_\rho(n)$ decreases with density, reaching $\mathcal{O}(0.4)$ of its saturation value at $n\sim 5\,n_0$. This indicates a preference against an excessively large high-density symmetry energy. While cooling inferences are model-dependent, such a suppression tends to reduce proton fractions in the deep core and can postpone direct-Urca thresholds to higher masses \cite{LattimerPrakash2016PhysRep}.

\subsection{Equation of state and the speed of sound}
\label{sec:results_eos}

\begin{figure*}[t]
  \centering
  \includegraphics[width=\textwidth]{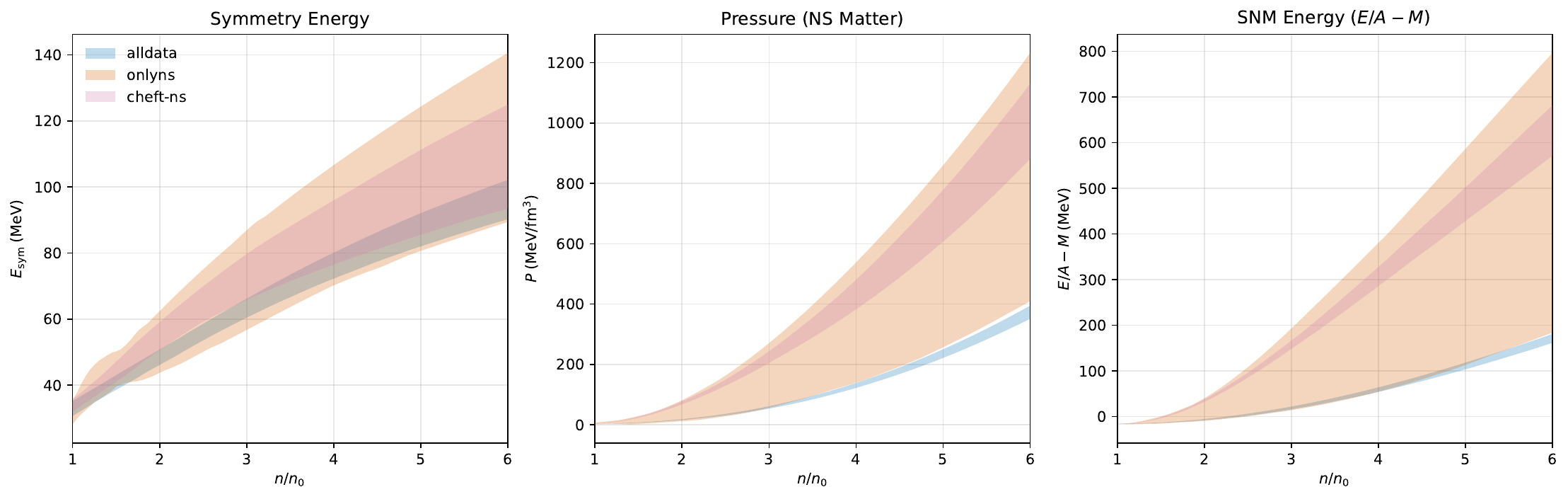}
  \caption{Comparison of reconstructed EOS quantities under different data selections. Shown are the symmetry energy $E_{\rm sym}(n)$ (left), the pressure of cold $\beta$-equilibrated matter $P(n)$ (middle), and the SNM binding energy $E/A - M$ (right), as functions of $n/n_0$. Colored bands represent 90\% credible intervals.}
  \label{fig:eos_comparison}
\end{figure*}

\begin{figure*}[t]
  \centering
  \includegraphics[width=\textwidth]{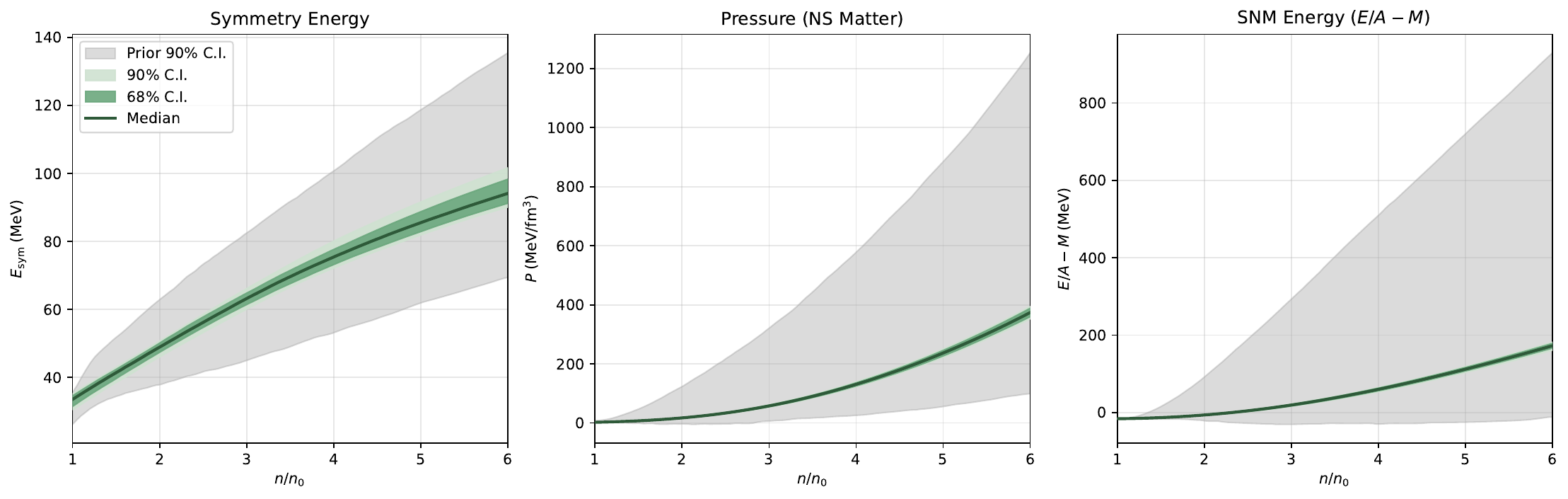}
  \caption{EOS quantities inferred from the fiducial \texttt{ALL} dataset. The gray band denotes the 90\% prior interval, while the dark and light green bands indicate the 68\% and 90\% posterior intervals, respectively; the solid curve is the posterior median.}
  \label{fig:eos_alldata}
\end{figure*}

\begin{figure}[t]
  \centering
  \includegraphics[width=\columnwidth]{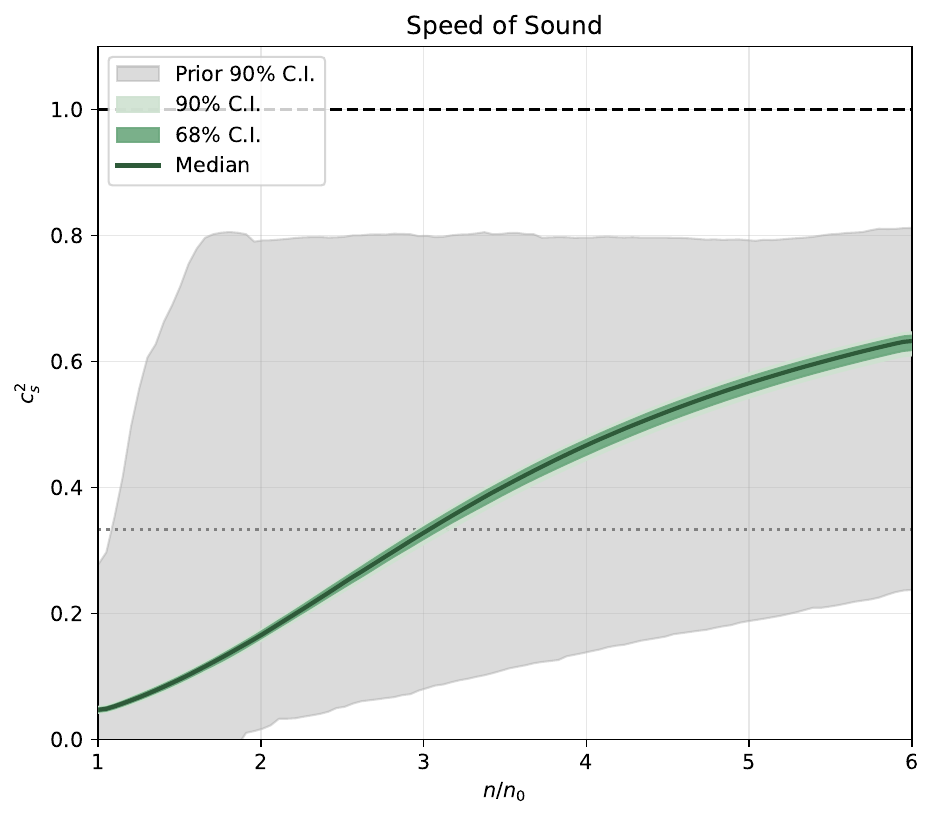}
  \caption{Squared speed of sound $c_s^2=dP/d\varepsilon$ for cold $\beta$-equilibrated matter inferred from \texttt{ALL}. The dashed line marks the causality bound $c_s^2=1$; the dotted line marks the conformal value $c_s^2=1/3$.}
  \label{fig:speed_of_sound}
\end{figure}

Translating the reconstructed couplings into thermodynamic observables, we examine the density dependence of the EOS. Figure~\ref{fig:eos_comparison} compares $E_{\rm sym}(n)$, the pressure of $\beta$-equilibrated matter, and the SNM energy across representative dataset selections. Figure~\ref{fig:eos_alldata} shows the corresponding posterior bands for \texttt{ALL} together with the prior envelope.

The comparison in Fig.~\ref{fig:eos_comparison} highlights the complementarity between terrestrial nuclear physics and astrophysical observations. NS-only constraints primarily inform the bulk $P(\varepsilon)$ relation but leave substantial freedom in the microscopic decomposition, leading to broader credible bands for quantities that depend sensitively on isoscalar/isovector separation (notably $E_{\rm sym}$). Adding $\chi$EFT anchors the EOS at sub- and near-saturation densities, tightening the symmetry-energy band and guiding the SNM energy. Combining all constraints yields a comparatively narrow EOS corridor over a broad density range: in the density regime relevant for heavy pulsars ($\sim 3$--$6\,n_0$), the \texttt{ALL} posterior favors pressures that remain compatible with intermediate-density HIC information while providing sufficient high-density stiffness to support $\sim 2\,M_\odot$ stars.

A stringent diagnostic is the squared speed of sound $c_s^2=dP/d\varepsilon$. In our nested sampling, samples violating causality ($c_s^2>1$) are rejected by construction. Figure~\ref{fig:speed_of_sound} shows that the posterior median increases with density, reaching $c_s^2\simeq 0.63$ at $n\approx 6\,n_0$, with relatively tight credible bands for \texttt{ALL}.

Figure~\ref{fig:speed_of_sound} also informs departures from conformal behavior. While weakly coupled ultrarelativistic matter approaches $c_s^2=1/3$ at asymptotically high densities \cite{Bedaque2015,Annala2020_NP}, our inference yields $c_s^2$ exceeding $1/3$ already at $n\approx 2.8\,n_0$. Such nonconformal stiffening is commonly observed in hadronic EOSs constrained by $\gtrsim 2\,M_\odot$ stars \cite{Tews2018ApJ,Reed2020PRC}. Because we do not impose the conformal bound a priori, the emergence of $c_s^2>1/3$ indicates that the matter realized in the density range probed by massive neutron-star cores is significantly nonconformal; this is in tension with a simple description that remains close to the conformal limit at a few times saturation density.

\subsection{Macroscopic observables: neutron-star structure}
\label{sec:results_macro}

\begin{figure*}[t]
  \centering
  \includegraphics[width=\textwidth]{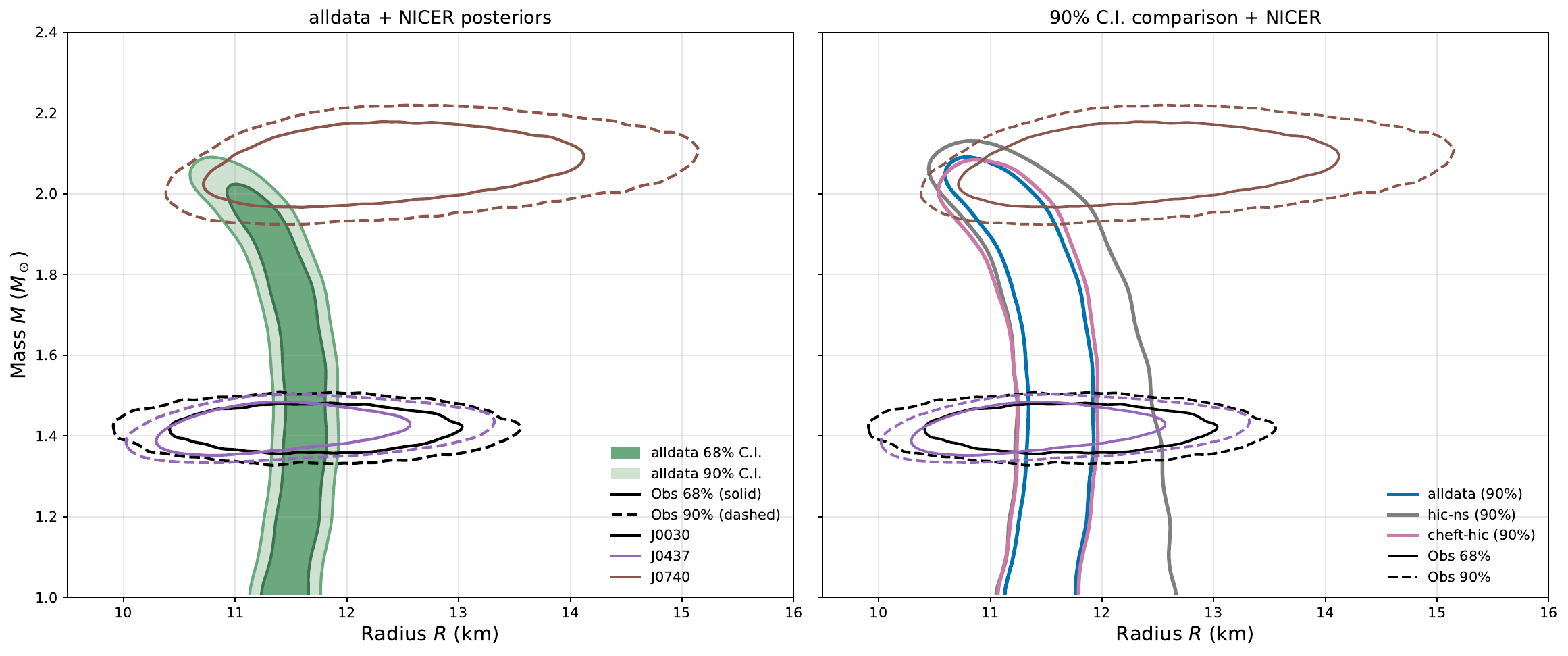}
  \caption{Mass--radius constraints from the inverse-mapped DD-RMF inference. \emph{Left:} 68\% (dark green) and 90\% (light green) credible regions for the \texttt{ALL} dataset. Overlaid contours show NICER posteriors for PSR~J0030$+$0451, PSR~J0437$-$4715, and PSR~J0740$+$6620; solid (dashed) lines denote 68\% (90\%) credible regions. \emph{Right:} Comparison of 90\% $M$--$R$ credible bands for \texttt{ALL}, \texttt{HIC+NS} (excluding $\chi$EFT), and \texttt{$\chi$EFT+HIC} (excluding NICER $M$--$R$ information).}
  \label{fig:mr_relations}
\end{figure*}

The reconstructed interaction is further tested through neutron-star structure, in particular the $M$--$R$ relation obtained from the TOV equations. Figure~\ref{fig:mr_relations} (left) shows the inferred $M$--$R$ credible regions for the fiducial \texttt{ALL} analysis, together with the NICER posteriors for PSR~J0030$+$0451 \cite{Miller2019J0030,Riley2019J0030}, PSR~J0437$-$4715 \cite{Choudhury2024J0437}, and PSR~J0740$+$6620 \cite{Riley2021J0740,Salmi2024J0740}.

\begin{figure*}[t]
  \centering
  \includegraphics[width=\textwidth]{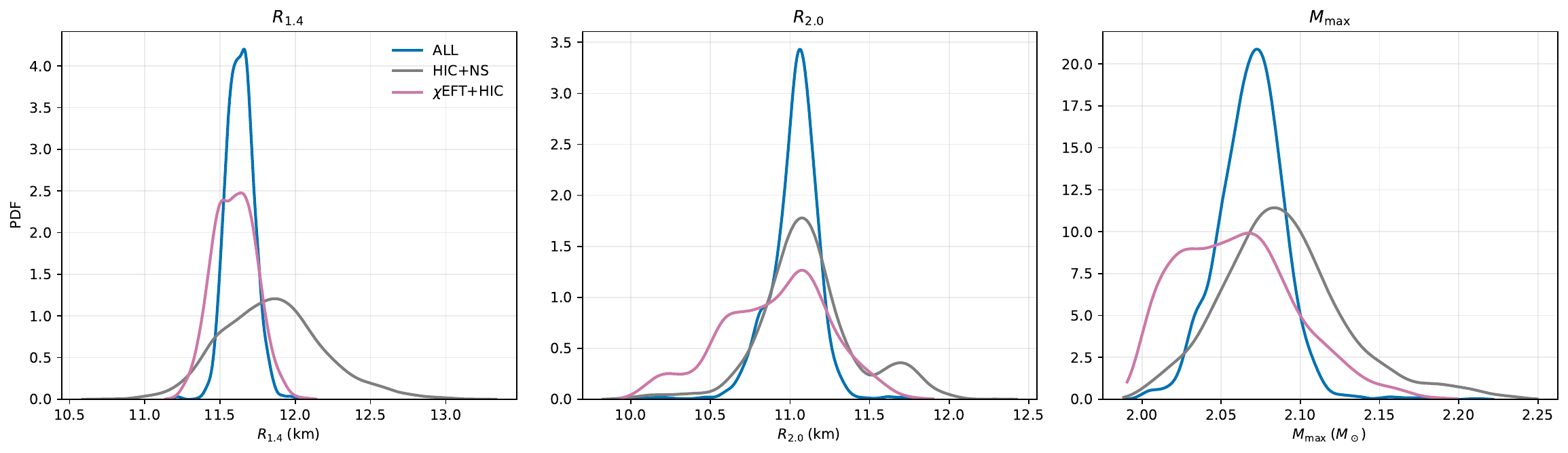}
  \caption{Marginalized posteriors of derived neutron-star observables: $R_{1.4}$ (left), $R_{2.0}$ (middle), and $M_{\max}$ (right). The comparison illustrates that low-density microscopic information predominantly constrains $R_{1.4}$, while astrophysical $M$--$R$ information sharpens the high-density sector governing $R_{2.0}$ and $M_{\max}$.}
  \label{fig:mr_pdfs}
\end{figure*}

To isolate the macroscopic impact of different constraint sectors, the right panel of Fig.~\ref{fig:mr_relations} compares the 90\% $M$--$R$ bands for three representative selections: \texttt{ALL}, \texttt{HIC+NS} (excluding $\chi$EFT), and the purely terrestrial \texttt{$\chi$EFT+HIC} (excluding NICER $M$--$R$ information). Figure~\ref{fig:mr_pdfs} shows the marginalized posteriors of derived quantities $R_{1.4}$, $R_{2.0}$, and $M_{\max}$, with numerical intervals summarized in Table~\ref{tab:mr_derived_ci}.

The canonical radius $R_{1.4}$ illustrates the importance of low-density neutron-rich information. The \texttt{HIC+NS} case yields a broad $R_{1.4}$ distribution (Fig.~\ref{fig:mr_pdfs}), reflecting residual degeneracies when the inference is not anchored by an \emph{ab initio} low-density band for neutron-rich matter. This is expected because the HIC band primarily constrains the isoscalar sector, whereas $R_{1.4}$ is strongly correlated with the symmetry energy around and below saturation. Including $\chi$EFT (\texttt{$\chi$EFT+HIC} and \texttt{ALL}) substantially sharpens the $R_{1.4}$ posterior. For \texttt{ALL} we obtain $R_{1.4} = 11.63^{+0.09}_{-0.08}\ {\rm km}$ (68\%~{\rm C.I.}) and $R_{1.4}=11.63^{+0.15}_{-0.13}\ {\rm km}$ (90\% C.I.) as listed in Table~\ref{tab:mr_derived_ci}. This compact canonical radius is consistent with the relatively soft symmetry-energy slope favored by $\chi$EFT and is compatible with tidal-deformability constraints from GW170817 \cite{Abbott2017GW170817,Abbott2018GW170817Radii}.

Massive-star properties probe the high-density stiffness. In the purely terrestrial \texttt{$\chi$EFT+HIC} inference, the high-density sector is only indirectly constrained through the imposed $M_{\max}$ penalty, allowing a softer-core tail that broadens the $M_{\max}$ and $R_{2.0}$ posteriors. Incorporating NICER $M$--$R$ posteriors---especially the high-mass target PSR~J0740$+$6620---suppresses this soft tail. Consequently, for \texttt{ALL} we find $M_{\max} = 2.070^{+0.017}_{-0.021}\,M_\odot$ (68\%~{\rm C.I.}) and $M_{\max}=2.070^{+0.030}_{-0.036}\,M_\odot$ (90\% C.I.), and $R_{2.0} = 11.05^{+0.11}_{-0.15}\ {\rm km}$ (68\%~{\rm C.I.}).
The proximity of $R_{2.0}$ to $R_{1.4}$ implies only a modest radius variation over a wide mass interval, consistent with an EOS that is comparatively soft at intermediate density (to satisfy $\chi$EFT/HIC) but stiffens at higher density to support heavy pulsars.

\begin{table*}[t]
\caption{Derived neutron-star macroscopic observables inferred from posterior EOS ensembles. We report 68\% and 90\% central credible intervals in the format $x^{+u}_{-l}$.}
\label{tab:mr_derived_ci}
\begin{ruledtabular}
\begin{tabular}{lcccc}
Data set & Observable & 68\% C.I. & 90\% C.I. \\
\hline
  \texttt{ALL} & $R_{1.4}$ (km) & $11.63^{+0.09}_{-0.08}$ & $11.63^{+0.15}_{-0.13}$ \\
  \texttt{ALL} & $R_{2.0}$ (km) & $11.05^{+0.11}_{-0.15}$ & $11.05^{+0.18}_{-0.28}$ \\
  \texttt{ALL} & $M_{\max}$ ($M_\odot$) & $2.070^{+0.017}_{-0.021}$ & $2.070^{+0.030}_{-0.036}$ \\
  \texttt{HIC+NS} & $R_{1.4}$ (km) & $11.84^{+0.33}_{-0.33}$ & $11.84^{+0.61}_{-0.50}$ \\
  \texttt{HIC+NS} & $R_{2.0}$ (km) & $11.09^{+0.28}_{-0.22}$ & $11.09^{+0.64}_{-0.42}$ \\
  \texttt{HIC+NS} & $M_{\max}$ ($M_\odot$) & $2.086^{+0.039}_{-0.034}$ & $2.086^{+0.077}_{-0.057}$ \\
  \texttt{$\chi$EFT+HIC} & $R_{1.4}$ (km) & $11.59^{+0.14}_{-0.14}$ & $11.59^{+0.23}_{-0.22}$ \\
  \texttt{$\chi$EFT+HIC} & $R_{2.0}$ (km) & $10.94^{+0.28}_{-0.37}$ & $10.94^{+0.48}_{-0.68}$ \\
  \texttt{$\chi$EFT+HIC} & $M_{\max}$ ($M_\odot$) & $2.058^{+0.038}_{-0.037}$ & $2.058^{+0.068}_{-0.051}$ \\
\end{tabular}
\end{ruledtabular}
\end{table*}

\subsection{Bayesian evidence and dataset compatibility}
\label{sec:results_evidence}

\begin{figure*}[t]
  \centering
  \includegraphics[width=\textwidth]{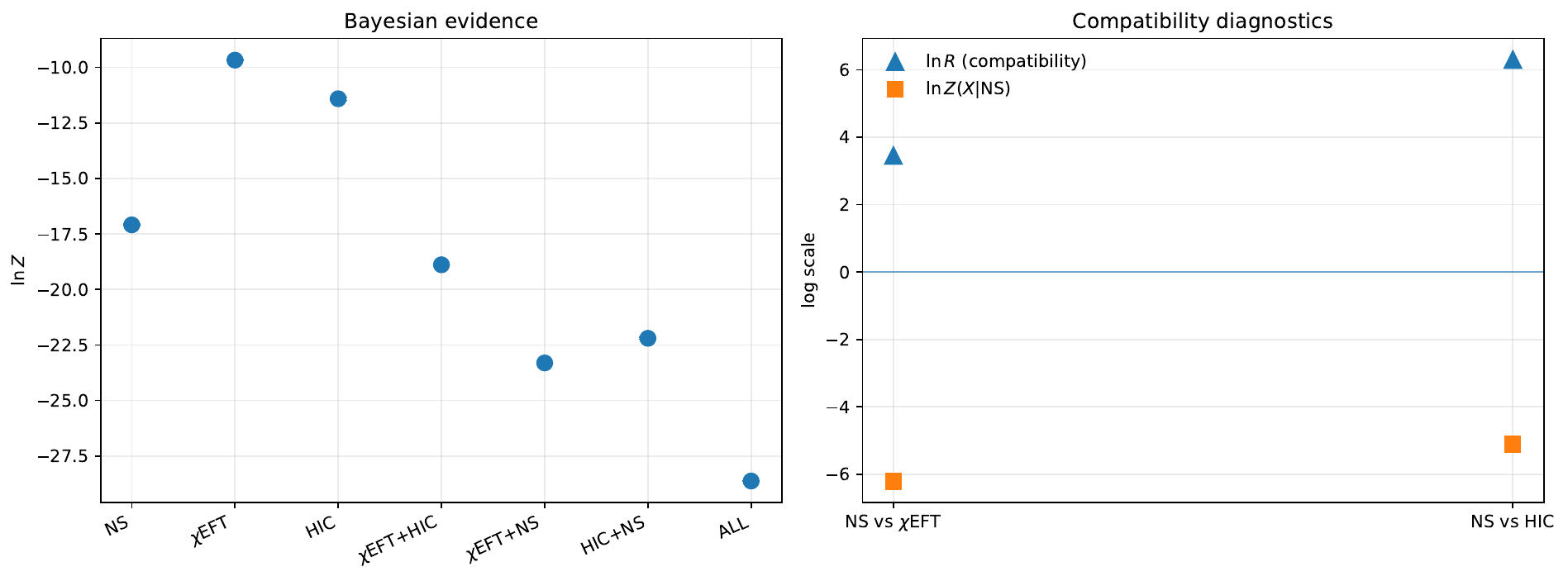}
  \caption{Bayesian evidence and dataset compatibility in the inverse-mapped DD-RMF model class. \emph{Left:} Global log-evidence $\ln\mathcal{Z}$ for each data selection; error bars denote numerical uncertainties from nested sampling. \emph{Right:} Compatibility diagnostics for NS paired with $\chi$EFT and HIC. Triangles show $\ln R_{12}=\ln Z(D_1,D_2)-\ln Z(D_1)-\ln Z(D_2)$; squares show the conditional evidence $\ln Z(X|{\rm NS})=\ln Z({\rm NS}+X)-\ln Z({\rm NS})$.}
  \label{fig:evidence_compat}
\end{figure*}

Nested sampling provides the marginal likelihood (Bayesian evidence) $\mathcal{Z}$, which quantifies the prior-predictive support for a given data selection within the assumed model class. The left panel of Fig.~\ref{fig:evidence_compat} shows $\ln\mathcal{Z}$ for the explored combinations. To assess incremental information gain when adding terrestrial constraints to NS-only inference, we report conditional evidences
\begin{equation}
\ln Z(X|{\rm NS}) \equiv \ln Z({\rm NS}+X)-\ln Z({\rm NS}),
\end{equation}
together with compatibility ratios
\begin{equation}
\ln R_{12}\equiv \ln Z(D_1,D_2)-\ln Z(D_1)-\ln Z(D_2),
\end{equation}
which compare the hypothesis that two datasets share the same parameters against the hypothesis that they are statistically independent \cite{KassRaftery1995}. Numerical values are summarized in Table~\ref{tab:evidence_compat}.

We find negative conditional evidences when augmenting NS-only inference with microscopic bands, e.g., $\ln Z(\chi{\rm EFT}|{\rm NS})=-6.22$ and $\ln Z({\rm HIC}|{\rm NS})=-5.11$. Such decreases are expected when additional constraints substantially compress the region of parameter space that retains non-negligible likelihood (an Occam-type reduction of prior volume). Because a negative $\ln Z(X|{\rm NS})$ may reflect both such compression and, in principle, dataset tension, it is most informative when interpreted together with $\ln R$.

The compatibility factors in Table~\ref{tab:evidence_compat} are positive for the relevant pairs, indicating that the different data sectors can be jointly accommodated within our inverse-mapped DD-RMF framework. In particular, we obtain $\ln R_{{\rm NS},\chi{\rm EFT}}=3.46$ and $\ln R_{{\rm NS},{\rm HIC}}=6.30$, corresponding to strong and very strong compatibility on the revised Jeffreys scale \cite{KassRaftery1995,Trotta2008BayesSky}. The purely terrestrial $\chi$EFT--HIC compatibility is also positive ($\ln R_{\chi{\rm EFT},{\rm HIC}}=2.20$), indicating non-negligible overlap after mapping through the same relativistic density functional and priors. The full \texttt{ALL} analysis yields a large multi-sector compatibility $\ln R_{{\rm NS},\chi{\rm EFT},{\rm HIC}}=9.55$, supporting the conclusion that, within the present model class and adopted priors, the combined microscopic and macroscopic constraints are statistically compatible.

\begin{table*}[t]
\caption{Bayesian evidence and compatibility summary for different data combinations under the inverse-mapped DD-RMF model. Global evidence is reported as $\ln Z \pm \sigma$. Relative evidence is $\Delta\ln Z \equiv \ln Z - \ln Z_{\rm NS}$. Conditional evidence terms quantify the incremental support of adding a dataset to the NS-only baseline, e.g., $\ln Z(\chi{\rm EFT}|{\rm NS})=\ln Z({\rm NS}+\chi{\rm EFT})-\ln Z({\rm NS})$. Dataset compatibility is quantified by $\ln R=\ln Z(D_1,D_2)-\ln Z(D_1)-\ln Z(D_2)$ (and its three-way analogue for \texttt{ALL}).}
\label{tab:evidence_compat}
\begin{ruledtabular}
\begin{tabular}{lcccc}
Data set $D$ & $\ln Z(D)$ & $\Delta\ln Z$ & Conditional evidence & Compatibility \\
\hline
  NS & $-17.09\pm0.03$ & $0.00$ & -- & -- \\
  $\chi$EFT & $-9.67\pm0.03$ & $7.42$ & -- & -- \\
  HIC & $-11.41\pm0.08$ & $5.68$ & -- & -- \\
  $\chi$EFT+HIC & $-18.89\pm0.02$ & $-1.80$ & -- & $2.20$ \\
  $\chi$EFT+NS & $-23.31\pm0.04$ & $-6.22$ & $-6.22$ & $3.46$ \\
  HIC+NS & $-22.20\pm0.04$ & $-5.11$ & $-5.11$ & $6.30$ \\
  ALL & $-28.62\pm0.02$ & $-11.53$ & $-11.53$ & $9.55$ \\
\end{tabular}
\end{ruledtabular}
\end{table*}

\section{Summary}
\label{sec:summary}

We developed a Bayesian inference framework for the EOS of cold, dense matter within a density-dependent relativistic mean-field (DD-RMF) model. A central feature is an explicit inverse-mapping procedure that connects a compact set of macroscopic empirical parameters to microscopic density-dependent interaction channels. Rather than sampling unconstrained phenomenological coefficients, we explore a physically interpretable 10-dimensional parameter space---including $K_0$, $M^*/M$, saturation properties, isovector parameters (e.g.\ $L$), and high-density asymptotic coupling limits---while enforcing thermodynamic consistency and applying explicit stability and causality filters. This construction enables a transparent diagnosis of how different density regimes and isospin sectors are constrained by different data sectors.

Conditioning the model on terrestrial nuclear physics ($\chi$EFT and HIC pressure bands) and astrophysical observations (NICER $M$--$R$ posteriors and heavy-pulsar mass information) highlights their complementarity. Adding $\chi$EFT substantially reduces isovector degeneracies and favors a relatively soft symmetry-energy slope, $L\simeq 38~\mathrm{MeV}$, which correlates with a compact canonical radius of $R_{1.4}=11.63^{+0.09}_{-0.08}~\mathrm{km}$ (68\% C.I.), consistent with tidal-deformability constraints from GW170817.

At suprasaturation densities, the combined dataset constrains the isoscalar sector strongly. To reproduce intermediate-density softness compatible with HIC information while maintaining sufficient high-density stiffness to support $\sim 2\,M_\odot$ neutron stars, the posterior favors a moderately large Dirac effective mass at saturation ($M^*/M\simeq 0.64$) together with correlated, non-vanishing high-density limits for the scalar and vector couplings. In the full \texttt{ALL} analysis, this implies $M_{\max}=2.070^{+0.017}_{-0.021}\,M_\odot$ and $R_{2.0}=11.05^{+0.11}_{-0.15}~\mathrm{km}$ (68\% C.I.).
The inferred sound-speed profile remains causal and exhibits pronounced nonconformal stiffening, with $c_s^2>1/3$ already at $n\approx 2.8\,n_0$, indicating that matter in the density range probed by massive neutron-star cores is significantly nonconformal.

Finally, evidence-based diagnostics indicate positive compatibility factors for the relevant dataset combinations; the full \texttt{ALL} analysis yields a large multi-sector compatibility ($\ln R \gtrsim 9$) within the present model class and adopted priors. Overall, our results suggest that, when endowed with sufficient microphysical flexibility and subjected to explicit consistency requirements, terrestrial nuclear constraints and multi-messenger neutron-star observations can be jointly accommodated, yielding a coherent and increasingly precise description of the neutron-star EOS. Future extensions of this framework will incorporate additional multi-messenger observables---including tidal deformabilities and post-merger information---and systematic studies of parametrization flexibility, including possible phase-transition dynamics at higher density.

\appendix

\section{Inverse mapping of the DD-RMF parameters}
\label{app:inverse_mapping}

In this appendix, we detail the inverse mapping procedure used to uniquely reconstruct the coupling strengths $G_{i0}$ and the density-dependence shape functions $f_i(n)$ from the 10-dimensional macroscopic parameter vector $\bm{\theta}$ defined in Eq.~\eqref{eq:theta_def}. The mapping is performed in four sequential steps.

\paragraph*{Step 1: Fix $G_{\sigma0}$ from $(M^*/M)_0$.}
At the saturation density $n=n_0$ in symmetric nuclear matter (SNM, $n_n=n_p=n_0/2$), we compute the Fermi momentum $k_{F0}$ and the scalar density $n_{s0}$ using Eq.~\eqref{eq:scalar_density} and the definition of the effective mass with $M_0^*=(M^*/M)_0\,M$. 
The scalar coupling strength at saturation is then given by
\begin{equation}
G_{\sigma0}=\frac{M-M_0^*}{n_{s0}}.
\label{eq:Gs0}
\end{equation}

\paragraph*{Step 2: Fix $G_{\omega0}$ from $E_0$ and the SNM energy at saturation.}
Using the kinetic energy density $\varepsilon_{N,0}$ at $(n_0/2,n_0/2)$, we enforce the energy per baryon in SNM to reproduce the binding energy $E_0$:
\begin{equation}
\frac{\varepsilon(n_0,0)}{n_0}=M+E_0
=\frac{\varepsilon_{N,0}}{n_0}+\frac12\,G_{\sigma0}\,\frac{n_{s0}^2}{n_0}+\frac12\,G_{\omega0}\,n_0.
\label{eq:solve_Gw0}
\end{equation}
This yields the vector coupling strength:
\begin{equation}
G_{\omega0}=\frac{2}{n_0}\left[(M+E_0)-\frac{\varepsilon_{N,0}}{n_0}-\frac12\,G_{\sigma0}\frac{n_{s0}^2}{n_0}\right].
\label{eq:Gw0}
\end{equation}

\paragraph*{Step 3: Fix $G_{\rho0}$ from $E_{\rm sym,0}$.}
Using the standard mean-field decomposition for the symmetry energy,
\begin{equation}
E_{\rm sym}(n)=E_{\rm sym}^{\rm kin}(n)+\frac12\,G_\rho(n)\,n,
\label{eq:Esym_decomp}
\end{equation}
where $E_{\rm sym}^{\rm kin}(n_0)=k_{F0}^2/(6E_{F0}^*)$, we obtain the isovector coupling strength:
\begin{equation}
G_{\rho0}=\frac{2}{n_0}\left[E_{\rm sym,0}-E_{\rm sym,0}^{\rm kin}\right].
\label{eq:Gr0}
\end{equation}

\paragraph*{Step 4: Determine the density-shape parameters in Eq.~\eqref{eq:tw_form}.}
We treat the isovector and isoscalar channels sequentially to ensure the stability of the inverse mapping.

\noindent
\textit{(i) Isovector channel $\rho$: matching $(L,K_{\rm sym})$.}
We write $G_\rho(n)=G_{\rho0} f_\rho^2(n)$ and define the first and second derivatives with respect to $x=n/n_0$ at saturation:
\begin{equation}
\eta_\rho \equiv \left.\frac{df_\rho}{dx}\right|_{x=1},\qquad
\kappa_\rho \equiv \left.\frac{d^2 f_\rho}{dx^2}\right|_{x=1}.
\label{eq:eta_kappa_def}
\end{equation}
Using Eq.~\eqref{eq:Esym_decomp} and the normalization $f_\rho(1)=1$, we find
\begin{align}
\left.\frac{dE_{\rm sym}}{dn}\right|_{n_0}
&=
\left.\frac{dE_{\rm sym}^{\rm kin}}{dn}\right|_{n_0}
+\frac12\,G_{\rho0}\left(1+2\eta_\rho\right),
\\
\left.\frac{d^2E_{\rm sym}}{dn^2}\right|_{n_0}
&=
\left.\frac{d^2E_{\rm sym}^{\rm kin}}{dn^2}\right|_{n_0}
+\frac{G_{\rho0}}{n_0}\left(2\eta_\rho+\eta_\rho^2+\kappa_\rho\right).
\end{align}
Inverting these relations gives the required derivatives:
\begin{align}
\eta_\rho &= \frac12\left[\frac{\displaystyle \frac{L}{3n_0}-\left.\frac{dE_{\rm sym}^{\rm kin}}{dn}\right|_{n_0}}
{\frac12\,G_{\rho0}} - 1\right],
\label{eq:eta_rho}\\
\kappa_\rho &=
\left(\frac{K_{\rm sym}}{9n_0^2}-\left.\frac{d^2E_{\rm sym}^{\rm kin}}{dn^2}\right|_{n_0}\right)\frac{n_0}{G_{\rho0}}
-\left(2\eta_\rho+\eta_\rho^2\right).
\label{eq:kappa_rho}
\end{align}
We then determine $(a_\rho,b_\rho,c_\rho,d_\rho)$ by solving the four constraints at $x=1$:
\begin{equation}
f_\rho(1)=1,\qquad
f_{\rho,\infty}=f_{\rho,\infty}^{\rm (input)},\qquad
\left.\frac{df_\rho}{dx}\right|_{1}=\eta_\rho,\qquad
\left.\frac{d^2 f_\rho}{dx^2}\right|_{1}=\kappa_\rho,
\label{eq:rho_constraints}
\end{equation}
which uniquely fixes the full isovector density dependence.

\noindent
\textit{(ii) Isoscalar channels $\sigma,\omega$: matching $(P(n_0)=0,\,K_0)$.}
To reduce the dimensionality and improve numerical stability, we set the shift parameter common to all channels, $d_\sigma=d_\omega=d_\rho\equiv d$, where $d_\rho$ is determined from the isovector sector. We enforce the asymptotic values $f_{\sigma,\infty}$ and $f_{\omega,\infty}$ through a two-parameter reduction:
\begin{equation}
f_i(x)=a_i\,\frac{1+b_i(x+d)^2}{1+c_i(x+d)^2},
\qquad
a_i = 1 + c_i(1+d)^2\,(1-f_{i,\infty}),
\qquad
b_i = \frac{f_{i,\infty}\,c_i}{a_i},
\label{eq:isoscalar_reduce}
\end{equation}
for $i \in \{\sigma, \omega\}$, so that $f_i(1)=1$ and $f_i(\infty)=f_{i,\infty}$ are satisfied identically. 
The remaining two unknowns $(c_\sigma,c_\omega)$ are determined numerically by enforcing the SNM thermodynamic conditions at saturation:
\begin{equation}
P(n_0,0)=0,
\qquad
K_0 = 9n_0\left.\frac{dP(n,0)}{dn}\right|_{n_0}.
\label{eq:K0_P0_constraints}
\end{equation}
In evaluating $P(n,0)$, the rearrangement contribution is explicitly included:
\begin{equation}
P(n,0)=P_N(n,0)+\frac12\,G_\omega(n)\,n^2-\frac12\,G_\sigma(n)\,n_s^2+n\,\Sigma_R(n),
\label{eq:P_snm}
\end{equation}
and $K_0$ is computed via a symmetric finite difference with a small step $\Delta n = 10^{-4}n_0$:
\begin{equation}
K_0 \approx 9n_0\,\frac{P(n_0+\Delta n,0)-P(n_0-\Delta n,0)}{2\Delta n}.
\label{eq:K0_fd}
\end{equation}
Equations \eqref{eq:K0_P0_constraints}--\eqref{eq:K0_fd} uniquely fix the isoscalar density dependence, completing the inverse mapping $\bm{\theta}\mapsto\{G_{i0},f_i(n)\}$.

\noindent{\bf Acknowledgement:} 
This work is supported by the National Natural Science Foundation of China (Grant No. 12275234) and the National SKA Program of China (Grant No. 2020SKA0120300).
%

\clearpage

\begin{thebibliography}{77}%
\makeatletter
\providecommand \@ifxundefined [1]{%
 \@ifx{#1\undefined}
}%
\providecommand \@ifnum [1]{%
 \ifnum #1\expandafter \@firstoftwo
 \else \expandafter \@secondoftwo
 \fi
}%
\providecommand \@ifx [1]{%
 \ifx #1\expandafter \@firstoftwo
 \else \expandafter \@secondoftwo
 \fi
}%
\providecommand \natexlab [1]{#1}%
\providecommand \enquote  [1]{``#1''}%
\providecommand \bibnamefont  [1]{#1}%
\providecommand \bibfnamefont [1]{#1}%
\providecommand \citenamefont [1]{#1}%
\providecommand \href@noop [0]{\@secondoftwo}%
\providecommand \href [0]{\begingroup \@sanitize@url \@href}%
\providecommand \@href[1]{\@@startlink{#1}\@@href}%
\providecommand \@@href[1]{\endgroup#1\@@endlink}%
\providecommand \@sanitize@url [0]{\catcode `\\12\catcode `\$12\catcode
  `\&12\catcode `\#12\catcode `\^12\catcode `\_12\catcode `\%12\relax}%
\providecommand \@@startlink[1]{}%
\providecommand \@@endlink[0]{}%
\providecommand \url  [0]{\begingroup\@sanitize@url \@url }%
\providecommand \@url [1]{\endgroup\@href {#1}{\urlprefix }}%
\providecommand \urlprefix  [0]{URL }%
\providecommand \Eprint [0]{\href }%
\providecommand \doibase [0]{http://dx.doi.org/}%
\providecommand \selectlanguage [0]{\@gobble}%
\providecommand \bibinfo  [0]{\@secondoftwo}%
\providecommand \bibfield  [0]{\@secondoftwo}%
\providecommand \translation [1]{[#1]}%
\providecommand \BibitemOpen [0]{}%
\providecommand \bibitemStop [0]{}%
\providecommand \bibitemNoStop [0]{.\EOS\space}%
\providecommand \EOS [0]{\spacefactor3000\relax}%
\providecommand \BibitemShut  [1]{\csname bibitem#1\endcsname}%
\let\auto@bib@innerbib\@empty
\bibitem [{\citenamefont {Oppenheimer}\ and\ \citenamefont
  {Volkoff}(1939)}]{OppenheimerVolkoff1939}%
  \BibitemOpen
  \bibfield  {author} {\bibinfo {author} {\bibfnamefont {J.~R.}\ \bibnamefont
  {Oppenheimer}}\ and\ \bibinfo {author} {\bibfnamefont {G.~M.}\ \bibnamefont
  {Volkoff}},\ }\href {\doibase 10.1103/PhysRev.55.374} {\bibfield  {journal}
  {\bibinfo  {journal} {Physical Review}\ }\textbf {\bibinfo {volume} {55}},\
  \bibinfo {pages} {374} (\bibinfo {year} {1939})}\BibitemShut {NoStop}%
\bibitem [{\citenamefont {Hinderer}(2008)}]{Hinderer2008}%
  \BibitemOpen
  \bibfield  {author} {\bibinfo {author} {\bibfnamefont {T.}~\bibnamefont
  {Hinderer}},\ }\href {\doibase 10.1086/533487} {\bibfield  {journal}
  {\bibinfo  {journal} {The Astrophysical Journal}\ }\textbf {\bibinfo {volume}
  {677}},\ \bibinfo {pages} {1216} (\bibinfo {year} {2008})},\ \Eprint
  {http://arxiv.org/abs/0711.2420} {arXiv:0711.2420} \BibitemShut {NoStop}%
\bibitem [{\citenamefont {Flanagan}\ and\ \citenamefont
  {Hinderer}(2008)}]{FlanaganHinderer2008}%
  \BibitemOpen
  \bibfield  {author} {\bibinfo {author} {\bibfnamefont {{\'E}.~{\'E}.}\
  \bibnamefont {Flanagan}}\ and\ \bibinfo {author} {\bibfnamefont
  {T.}~\bibnamefont {Hinderer}},\ }\href {\doibase 10.1103/PhysRevD.77.021502}
  {\bibfield  {journal} {\bibinfo  {journal} {Physical Review D}\ }\textbf
  {\bibinfo {volume} {77}},\ \bibinfo {pages} {021502} (\bibinfo {year}
  {2008})},\ \Eprint {http://arxiv.org/abs/0709.1915} {arXiv:0709.1915}
  \BibitemShut {NoStop}%
\bibitem [{\citenamefont {{\"O}zel}\ and\ \citenamefont
  {Freire}(2016)}]{OzelFreire2016ARAA}%
  \BibitemOpen
  \bibfield  {author} {\bibinfo {author} {\bibfnamefont {F.}~\bibnamefont
  {{\"O}zel}}\ and\ \bibinfo {author} {\bibfnamefont {P.}~\bibnamefont
  {Freire}},\ }\href {\doibase 10.1146/annurev-astro-081915-023322} {\bibfield
  {journal} {\bibinfo  {journal} {Annual Review of Astronomy and Astrophysics}\
  }\textbf {\bibinfo {volume} {54}},\ \bibinfo {pages} {401} (\bibinfo {year}
  {2016})}\BibitemShut {NoStop}%
\bibitem [{\citenamefont {Lattimer}\ and\ \citenamefont
  {Prakash}(2016)}]{LattimerPrakash2016PhysRep}%
  \BibitemOpen
  \bibfield  {author} {\bibinfo {author} {\bibfnamefont {J.~M.}\ \bibnamefont
  {Lattimer}}\ and\ \bibinfo {author} {\bibfnamefont {M.}~\bibnamefont
  {Prakash}},\ }\href {\doibase 10.1016/j.physrep.2015.12.005} {\bibfield
  {journal} {\bibinfo  {journal} {Physics Reports}\ }\textbf {\bibinfo {volume}
  {621}},\ \bibinfo {pages} {127} (\bibinfo {year} {2016})},\ \Eprint
  {http://arxiv.org/abs/1512.07820} {arXiv:1512.07820} \BibitemShut {NoStop}%
\bibitem [{\citenamefont {Oertel}\ \emph
  {et~al.}(2017{\natexlab{a}})\citenamefont {Oertel}, \citenamefont {Hempel},
  \citenamefont {Kl{\"a}hn},\ and\ \citenamefont {Typel}}]{Oertel2017RMP}%
  \BibitemOpen
  \bibfield  {author} {\bibinfo {author} {\bibfnamefont {M.}~\bibnamefont
  {Oertel}}, \bibinfo {author} {\bibfnamefont {M.}~\bibnamefont {Hempel}},
  \bibinfo {author} {\bibfnamefont {T.}~\bibnamefont {Kl{\"a}hn}}, \ and\
  \bibinfo {author} {\bibfnamefont {S.}~\bibnamefont {Typel}},\ }\href
  {\doibase 10.1103/RevModPhys.89.015007} {\bibfield  {journal} {\bibinfo
  {journal} {Reviews of Modern Physics}\ }\textbf {\bibinfo {volume} {89}},\
  \bibinfo {pages} {015007} (\bibinfo {year} {2017}{\natexlab{a}})}\BibitemShut
  {NoStop}%
\bibitem [{\citenamefont {Demorest}\ \emph {et~al.}(2010)\citenamefont
  {Demorest}, \citenamefont {Pennucci}, \citenamefont {Ransom}, \citenamefont
  {Roberts},\ and\ \citenamefont {Hessels}}]{Demorest2010Nature}%
  \BibitemOpen
  \bibfield  {author} {\bibinfo {author} {\bibfnamefont {P.~B.}\ \bibnamefont
  {Demorest}}, \bibinfo {author} {\bibfnamefont {T.}~\bibnamefont {Pennucci}},
  \bibinfo {author} {\bibfnamefont {S.}~\bibnamefont {Ransom}}, \bibinfo
  {author} {\bibfnamefont {M.}~\bibnamefont {Roberts}}, \ and\ \bibinfo
  {author} {\bibfnamefont {J.}~\bibnamefont {Hessels}},\ }\href {\doibase
  10.1038/nature09466} {\bibfield  {journal} {\bibinfo  {journal} {Nature}\
  }\textbf {\bibinfo {volume} {467}},\ \bibinfo {pages} {1081} (\bibinfo {year}
  {2010})},\ \Eprint {http://arxiv.org/abs/1010.5788} {arXiv:1010.5788}
  \BibitemShut {NoStop}%
\bibitem [{\citenamefont {Cromartie}\ \emph {et~al.}(2020)\citenamefont
  {Cromartie} \emph {et~al.}}]{Cromartie2020NatAstron}%
  \BibitemOpen
  \bibfield  {author} {\bibinfo {author} {\bibfnamefont {H.~T.}\ \bibnamefont
  {Cromartie}} \emph {et~al.},\ }\href {\doibase 10.1038/s41550-019-0880-2}
  {\bibfield  {journal} {\bibinfo  {journal} {Nature Astronomy}\ }\textbf
  {\bibinfo {volume} {4}},\ \bibinfo {pages} {72} (\bibinfo {year} {2020})},\
  \Eprint {http://arxiv.org/abs/1904.06759} {arXiv:1904.06759} \BibitemShut
  {NoStop}%
\bibitem [{\citenamefont {Fonseca}\ \emph {et~al.}(2021)\citenamefont {Fonseca}
  \emph {et~al.}}]{Fonseca2021ApJL}%
  \BibitemOpen
  \bibfield  {author} {\bibinfo {author} {\bibfnamefont {E.}~\bibnamefont
  {Fonseca}} \emph {et~al.},\ }\href@noop {} {\bibfield  {journal} {\bibinfo
  {journal} {The Astrophysical Journal Letters}\ }\textbf {\bibinfo {volume}
  {915}},\ \bibinfo {pages} {L12} (\bibinfo {year} {2021})}\BibitemShut
  {NoStop}%
\bibitem [{\citenamefont {Miller}\ \emph {et~al.}(2019)\citenamefont {Miller}
  \emph {et~al.}}]{Miller2019J0030}%
  \BibitemOpen
  \bibfield  {author} {\bibinfo {author} {\bibfnamefont {M.~C.}\ \bibnamefont
  {Miller}} \emph {et~al.},\ }\href {\doibase 10.3847/2041-8213/ab50c5}
  {\bibfield  {journal} {\bibinfo  {journal} {The Astrophysical Journal
  Letters}\ }\textbf {\bibinfo {volume} {887}},\ \bibinfo {pages} {L24}
  (\bibinfo {year} {2019})}\BibitemShut {NoStop}%
\bibitem [{\citenamefont {Riley}\ \emph {et~al.}(2019)\citenamefont {Riley}
  \emph {et~al.}}]{Riley2019J0030}%
  \BibitemOpen
  \bibfield  {author} {\bibinfo {author} {\bibfnamefont {T.~E.}\ \bibnamefont
  {Riley}} \emph {et~al.},\ }\href {\doibase 10.3847/2041-8213/ab481c}
  {\bibfield  {journal} {\bibinfo  {journal} {The Astrophysical Journal
  Letters}\ }\textbf {\bibinfo {volume} {887}},\ \bibinfo {pages} {L21}
  (\bibinfo {year} {2019})}\BibitemShut {NoStop}%
\bibitem [{\citenamefont {Riley}\ \emph {et~al.}(2021)\citenamefont {Riley}
  \emph {et~al.}}]{Riley2021J0740}%
  \BibitemOpen
  \bibfield  {author} {\bibinfo {author} {\bibfnamefont {T.~E.}\ \bibnamefont
  {Riley}} \emph {et~al.},\ }\href {\doibase 10.3847/2041-8213/ac0a81}
  {\bibfield  {journal} {\bibinfo  {journal} {The Astrophysical Journal
  Letters}\ }\textbf {\bibinfo {volume} {918}},\ \bibinfo {pages} {L27}
  (\bibinfo {year} {2021})},\ \Eprint {http://arxiv.org/abs/2105.06980}
  {arXiv:2105.06980} \BibitemShut {NoStop}%
\bibitem [{\citenamefont {Miller}\ \emph {et~al.}(2021)\citenamefont {Miller}
  \emph {et~al.}}]{Miller2021J0740}%
  \BibitemOpen
  \bibfield  {author} {\bibinfo {author} {\bibfnamefont {M.~C.}\ \bibnamefont
  {Miller}} \emph {et~al.},\ }\href {\doibase 10.3847/2041-8213/ac089b}
  {\bibfield  {journal} {\bibinfo  {journal} {The Astrophysical Journal
  Letters}\ }\textbf {\bibinfo {volume} {918}},\ \bibinfo {pages} {L28}
  (\bibinfo {year} {2021})},\ \Eprint {http://arxiv.org/abs/2105.06979}
  {arXiv:2105.06979} \BibitemShut {NoStop}%
\bibitem [{\citenamefont {Raaijmakers}\ \emph {et~al.}(2021)\citenamefont
  {Raaijmakers} \emph {et~al.}}]{Raaijmakers2021EOS}%
  \BibitemOpen
  \bibfield  {author} {\bibinfo {author} {\bibfnamefont {G.}~\bibnamefont
  {Raaijmakers}} \emph {et~al.},\ }\href {\doibase 10.3847/2041-8213/ac089a}
  {\bibfield  {journal} {\bibinfo  {journal} {The Astrophysical Journal
  Letters}\ }\textbf {\bibinfo {volume} {918}},\ \bibinfo {pages} {L29}
  (\bibinfo {year} {2021})},\ \Eprint {http://arxiv.org/abs/2105.06981}
  {arXiv:2105.06981} \BibitemShut {NoStop}%
\bibitem [{\citenamefont {Salmi}\ \emph {et~al.}(2022)\citenamefont {Salmi}
  \emph {et~al.}}]{Salmi2022J0740}%
  \BibitemOpen
  \bibfield  {author} {\bibinfo {author} {\bibfnamefont {T.}~\bibnamefont
  {Salmi}} \emph {et~al.},\ }\href {\doibase 10.3847/1538-4357/ac983d}
  {\bibfield  {journal} {\bibinfo  {journal} {The Astrophysical Journal}\
  }\textbf {\bibinfo {volume} {941}},\ \bibinfo {pages} {150} (\bibinfo {year}
  {2022})},\ \Eprint {http://arxiv.org/abs/2209.12840} {arXiv:2209.12840}
  \BibitemShut {NoStop}%
\bibitem [{\citenamefont {Salmi}\ \emph {et~al.}(2024)\citenamefont {Salmi}
  \emph {et~al.}}]{Salmi2024J0740}%
  \BibitemOpen
  \bibfield  {author} {\bibinfo {author} {\bibfnamefont {T.}~\bibnamefont
  {Salmi}} \emph {et~al.},\ }\href {\doibase 10.3847/1538-4357/ad5f1f}
  {\bibfield  {journal} {\bibinfo  {journal} {The Astrophysical Journal}\
  }\textbf {\bibinfo {volume} {974}},\ \bibinfo {pages} {294} (\bibinfo {year}
  {2024})}\BibitemShut {NoStop}%
\bibitem [{\citenamefont {Choudhury}\ \emph {et~al.}(2024)\citenamefont
  {Choudhury}, \citenamefont {Salmi}, \citenamefont {Vinciguerra},
  \citenamefont {Riley}, \citenamefont {Kini}, \citenamefont {Watts} \emph
  {et~al.}}]{Choudhury2024J0437}%
  \BibitemOpen
  \bibfield  {author} {\bibinfo {author} {\bibfnamefont {D.}~\bibnamefont
  {Choudhury}}, \bibinfo {author} {\bibfnamefont {T.}~\bibnamefont {Salmi}},
  \bibinfo {author} {\bibfnamefont {S.}~\bibnamefont {Vinciguerra}}, \bibinfo
  {author} {\bibfnamefont {T.~E.}\ \bibnamefont {Riley}}, \bibinfo {author}
  {\bibfnamefont {Y.}~\bibnamefont {Kini}}, \bibinfo {author} {\bibfnamefont
  {A.~L.}\ \bibnamefont {Watts}},  \emph {et~al.},\ }\href {\doibase
  10.3847/2041-8213/ad5a6f} {\bibfield  {journal} {\bibinfo  {journal} {The
  Astrophysical Journal Letters}\ }\textbf {\bibinfo {volume} {971}},\ \bibinfo
  {pages} {L20} (\bibinfo {year} {2024})},\ \Eprint
  {http://arxiv.org/abs/2407.06789} {arXiv:2407.06789} \BibitemShut {NoStop}%
\bibitem [{\citenamefont {Brandes}\ and\ \citenamefont
  {Weise}(2025)}]{BrandesWeise2025PRD}%
  \BibitemOpen
  \bibfield  {author} {\bibinfo {author} {\bibfnamefont {L.}~\bibnamefont
  {Brandes}}\ and\ \bibinfo {author} {\bibfnamefont {W.}~\bibnamefont
  {Weise}},\ }\href {\doibase 10.1103/PhysRevD.111.034005} {\bibfield
  {journal} {\bibinfo  {journal} {Physical Review D}\ }\textbf {\bibinfo
  {volume} {111}},\ \bibinfo {pages} {034005} (\bibinfo {year}
  {2025})}\BibitemShut {NoStop}%
\bibitem [{\citenamefont {Koehn}\ \emph {et~al.}(2025)\citenamefont {Koehn}
  \emph {et~al.}}]{Koehn2025PRX}%
  \BibitemOpen
  \bibfield  {author} {\bibinfo {author} {\bibfnamefont {M.}~\bibnamefont
  {Koehn}} \emph {et~al.},\ }\href {\doibase 10.1103/PhysRevX.15.021014}
  {\bibfield  {journal} {\bibinfo  {journal} {Physical Review X}\ }\textbf
  {\bibinfo {volume} {15}},\ \bibinfo {pages} {021014} (\bibinfo {year}
  {2025})}\BibitemShut {NoStop}%
\bibitem [{\citenamefont {Abbott}\ \emph {et~al.}(2017)\citenamefont {Abbott}
  \emph {et~al.}}]{Abbott2017GW170817}%
  \BibitemOpen
  \bibfield  {author} {\bibinfo {author} {\bibfnamefont {B.~P.}\ \bibnamefont
  {Abbott}} \emph {et~al.},\ }\href {\doibase 10.1103/PhysRevLett.119.161101}
  {\bibfield  {journal} {\bibinfo  {journal} {Physical Review Letters}\
  }\textbf {\bibinfo {volume} {119}},\ \bibinfo {pages} {161101} (\bibinfo
  {year} {2017})}\BibitemShut {NoStop}%
\bibitem [{\citenamefont {Abbott}\ \emph {et~al.}(2018)\citenamefont {Abbott}
  \emph {et~al.}}]{Abbott2018GW170817Radii}%
  \BibitemOpen
  \bibfield  {author} {\bibinfo {author} {\bibfnamefont {B.~P.}\ \bibnamefont
  {Abbott}} \emph {et~al.},\ }\href {\doibase 10.1103/PhysRevLett.121.161101}
  {\bibfield  {journal} {\bibinfo  {journal} {Physical Review Letters}\
  }\textbf {\bibinfo {volume} {121}},\ \bibinfo {pages} {161101} (\bibinfo
  {year} {2018})},\ \Eprint {http://arxiv.org/abs/1805.11581}
  {arXiv:1805.11581} \BibitemShut {NoStop}%
\bibitem [{\citenamefont {Abbott}\ \emph {et~al.}(2020)\citenamefont {Abbott}
  \emph {et~al.}}]{Abbott2020GW190425}%
  \BibitemOpen
  \bibfield  {author} {\bibinfo {author} {\bibfnamefont {B.~P.}\ \bibnamefont
  {Abbott}} \emph {et~al.},\ }\href {\doibase 10.3847/2041-8213/ab75f5}
  {\bibfield  {journal} {\bibinfo  {journal} {The Astrophysical Journal
  Letters}\ }\textbf {\bibinfo {volume} {892}},\ \bibinfo {pages} {L3}
  (\bibinfo {year} {2020})},\ \Eprint {http://arxiv.org/abs/2001.01761}
  {arXiv:2001.01761} \BibitemShut {NoStop}%
\bibitem [{\citenamefont {Abbott}\ \emph {et~al.}(2023)\citenamefont {Abbott}
  \emph {et~al.}}]{Abbott2023GWTC3}%
  \BibitemOpen
  \bibfield  {author} {\bibinfo {author} {\bibfnamefont {R.}~\bibnamefont
  {Abbott}} \emph {et~al.},\ }\href {\doibase 10.1103/PhysRevX.13.041039}
  {\bibfield  {journal} {\bibinfo  {journal} {Physical Review X}\ }\textbf
  {\bibinfo {volume} {13}},\ \bibinfo {pages} {041039} (\bibinfo {year}
  {2023})},\ \Eprint {http://arxiv.org/abs/2111.03606} {arXiv:2111.03606}
  \BibitemShut {NoStop}%
\bibitem [{\citenamefont {Bedaque}\ and\ \citenamefont
  {Steiner}(2015)}]{Bedaque2015}%
  \BibitemOpen
  \bibfield  {author} {\bibinfo {author} {\bibfnamefont {P.}~\bibnamefont
  {Bedaque}}\ and\ \bibinfo {author} {\bibfnamefont {A.~W.}\ \bibnamefont
  {Steiner}},\ }\href@noop {} {\bibfield  {journal} {\bibinfo  {journal}
  {Physical Review Letters}\ }\textbf {\bibinfo {volume} {114}},\ \bibinfo
  {pages} {031103} (\bibinfo {year} {2015})}\BibitemShut {NoStop}%
\bibitem [{\citenamefont {Annala}\ \emph
  {et~al.}(2020{\natexlab{a}})\citenamefont {Annala}, \citenamefont {Gorda},
  \citenamefont {Kurkela}, \citenamefont {N{\"a}ttil{\"a}},\ and\ \citenamefont
  {Vuorinen}}]{Annala2020}%
  \BibitemOpen
  \bibfield  {author} {\bibinfo {author} {\bibfnamefont {E.}~\bibnamefont
  {Annala}}, \bibinfo {author} {\bibfnamefont {T.}~\bibnamefont {Gorda}},
  \bibinfo {author} {\bibfnamefont {A.}~\bibnamefont {Kurkela}}, \bibinfo
  {author} {\bibfnamefont {J.}~\bibnamefont {N{\"a}ttil{\"a}}}, \ and\ \bibinfo
  {author} {\bibfnamefont {A.}~\bibnamefont {Vuorinen}},\ }\href@noop {}
  {\bibfield  {journal} {\bibinfo  {journal} {Nature Physics}\ }\textbf
  {\bibinfo {volume} {16}},\ \bibinfo {pages} {907} (\bibinfo {year}
  {2020}{\natexlab{a}})}\BibitemShut {NoStop}%
\bibitem [{\citenamefont {Altiparmak}\ \emph {et~al.}(2022)\citenamefont
  {Altiparmak}, \citenamefont {Ecker},\ and\ \citenamefont
  {Rezzolla}}]{Altiparmak2022}%
  \BibitemOpen
  \bibfield  {author} {\bibinfo {author} {\bibfnamefont {S.}~\bibnamefont
  {Altiparmak}}, \bibinfo {author} {\bibfnamefont {C.}~\bibnamefont {Ecker}}, \
  and\ \bibinfo {author} {\bibfnamefont {L.}~\bibnamefont {Rezzolla}},\
  }\href@noop {} {\bibfield  {journal} {\bibinfo  {journal} {Physical Review
  D}\ }\textbf {\bibinfo {volume} {106}},\ \bibinfo {pages} {083001} (\bibinfo
  {year} {2022})}\BibitemShut {NoStop}%
\bibitem [{\citenamefont {Hebeler}\ \emph {et~al.}(2010)\citenamefont
  {Hebeler}, \citenamefont {Lattimer}, \citenamefont {Pethick},\ and\
  \citenamefont {Schwenk}}]{Hebeler2010PRL}%
  \BibitemOpen
  \bibfield  {author} {\bibinfo {author} {\bibfnamefont {K.}~\bibnamefont
  {Hebeler}}, \bibinfo {author} {\bibfnamefont {J.~M.}\ \bibnamefont
  {Lattimer}}, \bibinfo {author} {\bibfnamefont {C.~J.}\ \bibnamefont
  {Pethick}}, \ and\ \bibinfo {author} {\bibfnamefont {A.}~\bibnamefont
  {Schwenk}},\ }\href {\doibase 10.1103/PhysRevLett.105.161102} {\bibfield
  {journal} {\bibinfo  {journal} {Physical Review Letters}\ }\textbf {\bibinfo
  {volume} {105}},\ \bibinfo {pages} {161102} (\bibinfo {year}
  {2010})}\BibitemShut {NoStop}%
\bibitem [{\citenamefont {Hebeler}\ \emph {et~al.}(2013)\citenamefont
  {Hebeler}, \citenamefont {Lattimer}, \citenamefont {Pethick},\ and\
  \citenamefont {Schwenk}}]{Hebeler2013ApJ}%
  \BibitemOpen
  \bibfield  {author} {\bibinfo {author} {\bibfnamefont {K.}~\bibnamefont
  {Hebeler}}, \bibinfo {author} {\bibfnamefont {J.~M.}\ \bibnamefont
  {Lattimer}}, \bibinfo {author} {\bibfnamefont {C.~J.}\ \bibnamefont
  {Pethick}}, \ and\ \bibinfo {author} {\bibfnamefont {A.}~\bibnamefont
  {Schwenk}},\ }\href {\doibase 10.1088/0004-637X/773/1/11} {\bibfield
  {journal} {\bibinfo  {journal} {The Astrophysical Journal}\ }\textbf
  {\bibinfo {volume} {773}},\ \bibinfo {pages} {11} (\bibinfo {year} {2013})},\
  \Eprint {http://arxiv.org/abs/1303.4662} {arXiv:1303.4662} \BibitemShut
  {NoStop}%
\bibitem [{\citenamefont {Drischler}\ \emph {et~al.}(2021)\citenamefont
  {Drischler}, \citenamefont {Holt},\ and\ \citenamefont
  {Wellenhofer}}]{DrischlerHoltWellenhofer2021ARNPS}%
  \BibitemOpen
  \bibfield  {author} {\bibinfo {author} {\bibfnamefont {C.}~\bibnamefont
  {Drischler}}, \bibinfo {author} {\bibfnamefont {J.~W.}\ \bibnamefont {Holt}},
  \ and\ \bibinfo {author} {\bibfnamefont {C.}~\bibnamefont {Wellenhofer}},\
  }\href {\doibase 10.1146/annurev-nucl-102419-041903} {\bibfield  {journal}
  {\bibinfo  {journal} {Annual Review of Nuclear and Particle Science}\
  }\textbf {\bibinfo {volume} {71}},\ \bibinfo {pages} {403} (\bibinfo {year}
  {2021})},\ \Eprint {http://arxiv.org/abs/2101.01709} {arXiv:2101.01709}
  \BibitemShut {NoStop}%
\bibitem [{\citenamefont {Drischler}\ \emph {et~al.}(2020)\citenamefont
  {Drischler}, \citenamefont {Melendez}, \citenamefont {Furnstahl},\ and\
  \citenamefont {Phillips}}]{Drischler2020PRC}%
  \BibitemOpen
  \bibfield  {author} {\bibinfo {author} {\bibfnamefont {C.}~\bibnamefont
  {Drischler}}, \bibinfo {author} {\bibfnamefont {J.~A.}\ \bibnamefont
  {Melendez}}, \bibinfo {author} {\bibfnamefont {R.~J.}\ \bibnamefont
  {Furnstahl}}, \ and\ \bibinfo {author} {\bibfnamefont {D.~R.}\ \bibnamefont
  {Phillips}},\ }\href {\doibase 10.1103/PhysRevC.102.054315} {\bibfield
  {journal} {\bibinfo  {journal} {Physical Review C}\ }\textbf {\bibinfo
  {volume} {102}},\ \bibinfo {pages} {054315} (\bibinfo {year} {2020})},\
  \Eprint {http://arxiv.org/abs/2004.07805} {arXiv:2004.07805} \BibitemShut
  {NoStop}%
\bibitem [{\citenamefont {Keller}\ \emph {et~al.}(2023)\citenamefont {Keller}
  \emph {et~al.}}]{Keller2023PRL}%
  \BibitemOpen
  \bibfield  {author} {\bibinfo {author} {\bibfnamefont {J.}~\bibnamefont
  {Keller}} \emph {et~al.},\ }\href {\doibase 10.1103/PhysRevLett.130.072701}
  {\bibfield  {journal} {\bibinfo  {journal} {Physical Review Letters}\
  }\textbf {\bibinfo {volume} {130}},\ \bibinfo {pages} {072701} (\bibinfo
  {year} {2023})}\BibitemShut {NoStop}%
\bibitem [{\citenamefont {Lim}\ and\ \citenamefont
  {Schwenk}(2024)}]{LimSchwenk2024PRC}%
  \BibitemOpen
  \bibfield  {author} {\bibinfo {author} {\bibfnamefont {Y.}~\bibnamefont
  {Lim}}\ and\ \bibinfo {author} {\bibfnamefont {A.}~\bibnamefont {Schwenk}},\
  }\href {\doibase 10.1103/PhysRevC.109.035801} {\bibfield  {journal} {\bibinfo
   {journal} {Physical Review C}\ }\textbf {\bibinfo {volume} {109}},\ \bibinfo
  {pages} {035801} (\bibinfo {year} {2024})},\ \Eprint
  {http://arxiv.org/abs/2307.04063} {arXiv:2307.04063} \BibitemShut {NoStop}%
\bibitem [{\citenamefont {Rutherford}\ \emph {et~al.}(2024)\citenamefont
  {Rutherford}, \citenamefont {Mendes}, \citenamefont {Svensson}, \citenamefont
  {Schwenk}, \citenamefont {Watts}, \citenamefont {Hebeler}, \citenamefont
  {Keller}, \citenamefont {Prescod-Weinstein}, \citenamefont {Choudhury},
  \citenamefont {Raaijmakers}, \citenamefont {Salmi}, \citenamefont
  {Timmerman}, \citenamefont {Vinciguerra}, \citenamefont {Guillot},\ and\
  \citenamefont {Lattimer}}]{Rutherford_2024}%
  \BibitemOpen
  \bibfield  {author} {\bibinfo {author} {\bibfnamefont {N.}~\bibnamefont
  {Rutherford}}, \bibinfo {author} {\bibfnamefont {M.}~\bibnamefont {Mendes}},
  \bibinfo {author} {\bibfnamefont {I.}~\bibnamefont {Svensson}}, \bibinfo
  {author} {\bibfnamefont {A.}~\bibnamefont {Schwenk}}, \bibinfo {author}
  {\bibfnamefont {A.~L.}\ \bibnamefont {Watts}}, \bibinfo {author}
  {\bibfnamefont {K.}~\bibnamefont {Hebeler}}, \bibinfo {author} {\bibfnamefont
  {J.}~\bibnamefont {Keller}}, \bibinfo {author} {\bibfnamefont
  {C.}~\bibnamefont {Prescod-Weinstein}}, \bibinfo {author} {\bibfnamefont
  {D.}~\bibnamefont {Choudhury}}, \bibinfo {author} {\bibfnamefont
  {G.}~\bibnamefont {Raaijmakers}}, \bibinfo {author} {\bibfnamefont
  {T.}~\bibnamefont {Salmi}}, \bibinfo {author} {\bibfnamefont
  {P.}~\bibnamefont {Timmerman}}, \bibinfo {author} {\bibfnamefont
  {S.}~\bibnamefont {Vinciguerra}}, \bibinfo {author} {\bibfnamefont
  {S.}~\bibnamefont {Guillot}}, \ and\ \bibinfo {author} {\bibfnamefont
  {J.~M.}\ \bibnamefont {Lattimer}},\ }\href {\doibase
  10.3847/2041-8213/ad5f02} {\bibfield  {journal} {\bibinfo  {journal} {The
  Astrophysical Journal Letters}\ }\textbf {\bibinfo {volume} {971}},\ \bibinfo
  {pages} {L19} (\bibinfo {year} {2024})}\BibitemShut {NoStop}%
\bibitem [{\citenamefont {Danielewicz}\ \emph
  {et~al.}(2002{\natexlab{a}})\citenamefont {Danielewicz}, \citenamefont
  {Lacey},\ and\ \citenamefont {Lynch}}]{Danielewicz2002Science}%
  \BibitemOpen
  \bibfield  {author} {\bibinfo {author} {\bibfnamefont {P.}~\bibnamefont
  {Danielewicz}}, \bibinfo {author} {\bibfnamefont {R.}~\bibnamefont {Lacey}},
  \ and\ \bibinfo {author} {\bibfnamefont {W.~G.}\ \bibnamefont {Lynch}},\
  }\href {\doibase 10.1126/science.1078070} {\bibfield  {journal} {\bibinfo
  {journal} {Science}\ }\textbf {\bibinfo {volume} {298}},\ \bibinfo {pages}
  {1592} (\bibinfo {year} {2002}{\natexlab{a}})},\ \Eprint
  {http://arxiv.org/abs/nucl-th/0208016} {arXiv:nucl-th/0208016} \BibitemShut
  {NoStop}%
\bibitem [{\citenamefont {Russotto}\ \emph {et~al.}(2016)\citenamefont
  {Russotto} \emph {et~al.}}]{Russotto2016ASYEOS}%
  \BibitemOpen
  \bibfield  {author} {\bibinfo {author} {\bibfnamefont {P.}~\bibnamefont
  {Russotto}} \emph {et~al.},\ }\href {\doibase 10.1103/PhysRevC.94.034608}
  {\bibfield  {journal} {\bibinfo  {journal} {Physical Review C}\ }\textbf
  {\bibinfo {volume} {94}},\ \bibinfo {pages} {034608} (\bibinfo {year}
  {2016})},\ \Eprint {http://arxiv.org/abs/1608.04332} {arXiv:1608.04332}
  \BibitemShut {NoStop}%
\bibitem [{\citenamefont {S{\o}rensen}\ \emph {et~al.}(2024)\citenamefont
  {S{\o}rensen} \emph {et~al.}}]{Sorensen2024PPNP}%
  \BibitemOpen
  \bibfield  {author} {\bibinfo {author} {\bibfnamefont {A.}~\bibnamefont
  {S{\o}rensen}} \emph {et~al.},\ }\href {\doibase 10.1016/j.ppnp.2023.104080}
  {\bibfield  {journal} {\bibinfo  {journal} {Progress in Particle and Nuclear
  Physics}\ }\textbf {\bibinfo {volume} {134}},\ \bibinfo {pages} {104080}
  (\bibinfo {year} {2024})}\BibitemShut {NoStop}%
\bibitem [{\citenamefont {Huth}\ \emph {et~al.}(2022)\citenamefont {Huth} \emph
  {et~al.}}]{Huth2022Nature}%
  \BibitemOpen
  \bibfield  {author} {\bibinfo {author} {\bibfnamefont {S.}~\bibnamefont
  {Huth}} \emph {et~al.},\ }\href {\doibase 10.1038/s41586-022-04750-w}
  {\bibfield  {journal} {\bibinfo  {journal} {Nature}\ }\textbf {\bibinfo
  {volume} {606}},\ \bibinfo {pages} {276} (\bibinfo {year}
  {2022})}\BibitemShut {NoStop}%
\bibitem [{\citenamefont {Adhikari}\ \emph {et~al.}(2021)\citenamefont
  {Adhikari} \emph {et~al.}}]{Adhikari2021PREX2}%
  \BibitemOpen
  \bibfield  {author} {\bibinfo {author} {\bibfnamefont {D.}~\bibnamefont
  {Adhikari}} \emph {et~al.},\ }\href {\doibase 10.1103/PhysRevLett.126.172502}
  {\bibfield  {journal} {\bibinfo  {journal} {Physical Review Letters}\
  }\textbf {\bibinfo {volume} {126}},\ \bibinfo {pages} {172502} (\bibinfo
  {year} {2021})},\ \Eprint {http://arxiv.org/abs/2102.10767}
  {arXiv:2102.10767} \BibitemShut {NoStop}%
\bibitem [{\citenamefont {Adhikari}\ \emph {et~al.}(2022)\citenamefont
  {Adhikari} \emph {et~al.}}]{Adhikari2022CREX}%
  \BibitemOpen
  \bibfield  {author} {\bibinfo {author} {\bibfnamefont {D.}~\bibnamefont
  {Adhikari}} \emph {et~al.},\ }\href {\doibase 10.1103/PhysRevLett.129.042501}
  {\bibfield  {journal} {\bibinfo  {journal} {Physical Review Letters}\
  }\textbf {\bibinfo {volume} {129}},\ \bibinfo {pages} {042501} (\bibinfo
  {year} {2022})},\ \Eprint {http://arxiv.org/abs/2205.11593}
  {arXiv:2205.11593} \BibitemShut {NoStop}%
\bibitem [{\citenamefont {Walecka}(1974)}]{Walecka1974}%
  \BibitemOpen
  \bibfield  {author} {\bibinfo {author} {\bibfnamefont {J.~D.}\ \bibnamefont
  {Walecka}},\ }\href {\doibase 10.1016/0003-4916(74)90208-5} {\bibfield
  {journal} {\bibinfo  {journal} {Annals of Physics}\ }\textbf {\bibinfo
  {volume} {83}},\ \bibinfo {pages} {491} (\bibinfo {year} {1974})}\BibitemShut
  {NoStop}%
\bibitem [{\citenamefont {Boguta}\ and\ \citenamefont
  {Bodmer}(1977)}]{BogutaBodmer1977}%
  \BibitemOpen
  \bibfield  {author} {\bibinfo {author} {\bibfnamefont {J.}~\bibnamefont
  {Boguta}}\ and\ \bibinfo {author} {\bibfnamefont {A.~R.}\ \bibnamefont
  {Bodmer}},\ }\href {\doibase 10.1016/0375-9474(77)90626-1} {\bibfield
  {journal} {\bibinfo  {journal} {Nuclear Physics A}\ }\textbf {\bibinfo
  {volume} {292}},\ \bibinfo {pages} {413} (\bibinfo {year}
  {1977})}\BibitemShut {NoStop}%
\bibitem [{\citenamefont {Ring}(1996)}]{Ring1996}%
  \BibitemOpen
  \bibfield  {author} {\bibinfo {author} {\bibfnamefont {P.}~\bibnamefont
  {Ring}},\ }\href {\doibase 10.1016/0146-6410(96)00054-3} {\bibfield
  {journal} {\bibinfo  {journal} {Progress in Particle and Nuclear Physics}\
  }\textbf {\bibinfo {volume} {37}},\ \bibinfo {pages} {193} (\bibinfo {year}
  {1996})}\BibitemShut {NoStop}%
\bibitem [{\citenamefont {Serot}\ and\ \citenamefont
  {Walecka}(1997)}]{SerotWalecka1997}%
  \BibitemOpen
  \bibfield  {author} {\bibinfo {author} {\bibfnamefont {B.~D.}\ \bibnamefont
  {Serot}}\ and\ \bibinfo {author} {\bibfnamefont {J.~D.}\ \bibnamefont
  {Walecka}},\ }\href {\doibase 10.1142/S0218301397000299} {\bibfield
  {journal} {\bibinfo  {journal} {International Journal of Modern Physics E}\
  }\textbf {\bibinfo {volume} {6}},\ \bibinfo {pages} {515} (\bibinfo {year}
  {1997})},\ \Eprint {http://arxiv.org/abs/nucl-th/9701058}
  {arXiv:nucl-th/9701058} \BibitemShut {NoStop}%
\bibitem [{\citenamefont {Lalazissis}\ \emph {et~al.}(1997)\citenamefont
  {Lalazissis}, \citenamefont {K{\"o}nig},\ and\ \citenamefont
  {Ring}}]{Lalazissis1997NL3}%
  \BibitemOpen
  \bibfield  {author} {\bibinfo {author} {\bibfnamefont {G.~A.}\ \bibnamefont
  {Lalazissis}}, \bibinfo {author} {\bibfnamefont {J.}~\bibnamefont
  {K{\"o}nig}}, \ and\ \bibinfo {author} {\bibfnamefont {P.}~\bibnamefont
  {Ring}},\ }\href {\doibase 10.1103/PhysRevC.55.540} {\bibfield  {journal}
  {\bibinfo  {journal} {Physical Review C}\ }\textbf {\bibinfo {volume} {55}},\
  \bibinfo {pages} {540} (\bibinfo {year} {1997})}\BibitemShut {NoStop}%
\bibitem [{\citenamefont {Todd-Rutel}\ and\ \citenamefont
  {Piekarewicz}(2005)}]{ToddRutelPiekarewicz2005FSUGold}%
  \BibitemOpen
  \bibfield  {author} {\bibinfo {author} {\bibfnamefont {B.~G.}\ \bibnamefont
  {Todd-Rutel}}\ and\ \bibinfo {author} {\bibfnamefont {J.}~\bibnamefont
  {Piekarewicz}},\ }\href {\doibase 10.1103/PhysRevLett.95.122501} {\bibfield
  {journal} {\bibinfo  {journal} {Physical Review Letters}\ }\textbf {\bibinfo
  {volume} {95}},\ \bibinfo {pages} {122501} (\bibinfo {year} {2005})},\
  \Eprint {http://arxiv.org/abs/nucl-th/0504034} {arXiv:nucl-th/0504034}
  \BibitemShut {NoStop}%
\bibitem [{\citenamefont {Typel}\ and\ \citenamefont
  {Wolter}(1999)}]{TypelWolter1999}%
  \BibitemOpen
  \bibfield  {author} {\bibinfo {author} {\bibfnamefont {S.}~\bibnamefont
  {Typel}}\ and\ \bibinfo {author} {\bibfnamefont {H.~H.}\ \bibnamefont
  {Wolter}},\ }\href {\doibase 10.1016/S0375-9474(99)00310-3} {\bibfield
  {journal} {\bibinfo  {journal} {Nuclear Physics A}\ }\textbf {\bibinfo
  {volume} {656}},\ \bibinfo {pages} {331} (\bibinfo {year}
  {1999})}\BibitemShut {NoStop}%
\bibitem [{\citenamefont {Lalazissis}\ \emph {et~al.}(2005)\citenamefont
  {Lalazissis}, \citenamefont {Nik{\v{s}}i{\'c}}, \citenamefont {Vretenar},\
  and\ \citenamefont {Ring}}]{Lalazissis2005DDME2}%
  \BibitemOpen
  \bibfield  {author} {\bibinfo {author} {\bibfnamefont {G.~A.}\ \bibnamefont
  {Lalazissis}}, \bibinfo {author} {\bibfnamefont {T.}~\bibnamefont
  {Nik{\v{s}}i{\'c}}}, \bibinfo {author} {\bibfnamefont {D.}~\bibnamefont
  {Vretenar}}, \ and\ \bibinfo {author} {\bibfnamefont {P.}~\bibnamefont
  {Ring}},\ }\href {\doibase 10.1103/PhysRevC.71.024312} {\bibfield  {journal}
  {\bibinfo  {journal} {Physical Review C}\ }\textbf {\bibinfo {volume} {71}},\
  \bibinfo {pages} {024312} (\bibinfo {year} {2005})}\BibitemShut {NoStop}%
\bibitem [{\citenamefont {Nik{\v{s}}i{\'c}}\ \emph {et~al.}(2008)\citenamefont
  {Nik{\v{s}}i{\'c}}, \citenamefont {Vretenar},\ and\ \citenamefont
  {Ring}}]{Niksic2008DDPC1}%
  \BibitemOpen
  \bibfield  {author} {\bibinfo {author} {\bibfnamefont {T.}~\bibnamefont
  {Nik{\v{s}}i{\'c}}}, \bibinfo {author} {\bibfnamefont {D.}~\bibnamefont
  {Vretenar}}, \ and\ \bibinfo {author} {\bibfnamefont {P.}~\bibnamefont
  {Ring}},\ }\href {\doibase 10.1103/PhysRevC.78.034318} {\bibfield  {journal}
  {\bibinfo  {journal} {Physical Review C}\ }\textbf {\bibinfo {volume} {78}},\
  \bibinfo {pages} {034318} (\bibinfo {year} {2008})}\BibitemShut {NoStop}%
\bibitem [{\citenamefont {Skilling}(2004)}]{Skilling2004NestedSampling}%
  \BibitemOpen
  \bibfield  {author} {\bibinfo {author} {\bibfnamefont {J.}~\bibnamefont
  {Skilling}},\ }in\ \href {\doibase 10.1063/1.1835238} {\emph {\bibinfo
  {booktitle} {AIP Conference Proceedings}}},\ Vol.\ \bibinfo {volume} {735}\
  (\bibinfo {year} {2004})\ pp.\ \bibinfo {pages} {395--405}\BibitemShut
  {NoStop}%
\bibitem [{\citenamefont {Feroz}\ \emph {et~al.}(2009)\citenamefont {Feroz},
  \citenamefont {Hobson},\ and\ \citenamefont {Bridges}}]{Feroz2009MultiNest}%
  \BibitemOpen
  \bibfield  {author} {\bibinfo {author} {\bibfnamefont {F.}~\bibnamefont
  {Feroz}}, \bibinfo {author} {\bibfnamefont {M.~P.}\ \bibnamefont {Hobson}}, \
  and\ \bibinfo {author} {\bibfnamefont {M.}~\bibnamefont {Bridges}},\ }\href
  {\doibase 10.1111/j.1365-2966.2009.14548.x} {\bibfield  {journal} {\bibinfo
  {journal} {Monthly Notices of the Royal Astronomical Society}\ }\textbf
  {\bibinfo {volume} {398}},\ \bibinfo {pages} {1601} (\bibinfo {year}
  {2009})}\BibitemShut {NoStop}%
\bibitem [{\citenamefont {Malik}\ \emph {et~al.}(2022)\citenamefont {Malik},
  \citenamefont {Ferreira}, \citenamefont {Agrawal},\ and\ \citenamefont
  {Provid{\^e}ncia}}]{Malik2022_ApJ930-17}%
  \BibitemOpen
  \bibfield  {author} {\bibinfo {author} {\bibfnamefont {T.}~\bibnamefont
  {Malik}}, \bibinfo {author} {\bibfnamefont {M.}~\bibnamefont {Ferreira}},
  \bibinfo {author} {\bibfnamefont {B.~K.}\ \bibnamefont {Agrawal}}, \ and\
  \bibinfo {author} {\bibfnamefont {C.}~\bibnamefont {Provid{\^e}ncia}},\
  }\href {\doibase 3} {\bibfield  {journal} {\bibinfo  {journal} {Astrophys.
  J.}\ }\textbf {\bibinfo {volume} {930}},\ \bibinfo {pages} {17} (\bibinfo
  {year} {2022})}\BibitemShut {NoStop}%
\bibitem [{\citenamefont {Traversi}\ \emph {et~al.}(2020)\citenamefont
  {Traversi}, \citenamefont {Char},\ and\ \citenamefont
  {Pagliara}}]{Traversi2020PRC}%
  \BibitemOpen
  \bibfield  {author} {\bibinfo {author} {\bibfnamefont {S.}~\bibnamefont
  {Traversi}}, \bibinfo {author} {\bibfnamefont {P.}~\bibnamefont {Char}}, \
  and\ \bibinfo {author} {\bibfnamefont {G.}~\bibnamefont {Pagliara}},\
  }\href@noop {} {\bibfield  {journal} {\bibinfo  {journal} {Physical Review
  C}\ }\textbf {\bibinfo {volume} {102}},\ \bibinfo {pages} {024302} (\bibinfo
  {year} {2020})}\BibitemShut {NoStop}%
\bibitem [{\citenamefont {Typel}(2005)}]{Typel2005PRC}%
  \BibitemOpen
  \bibfield  {author} {\bibinfo {author} {\bibfnamefont {S.}~\bibnamefont
  {Typel}},\ }\href {\doibase 10.1103/PhysRevC.71.064301} {\bibfield  {journal}
  {\bibinfo  {journal} {Phys. Rev. C}\ }\textbf {\bibinfo {volume} {71}},\
  \bibinfo {pages} {064301} (\bibinfo {year} {2005})},\ \Eprint
  {http://arxiv.org/abs/nucl-th/0501056} {arXiv:nucl-th/0501056} \BibitemShut
  {NoStop}%
\bibitem [{\citenamefont {Xia}\ \emph {et~al.}(2024)\citenamefont {Xia},
  \citenamefont {Xie},\ and\ \citenamefont {Bakhiet}}]{Xia2024_PRD110-114009}%
  \BibitemOpen
  \bibfield  {author} {\bibinfo {author} {\bibfnamefont {C.-J.}\ \bibnamefont
  {Xia}}, \bibinfo {author} {\bibfnamefont {W.-J.}\ \bibnamefont {Xie}}, \ and\
  \bibinfo {author} {\bibfnamefont {M.}~\bibnamefont {Bakhiet}},\ }\href
  {\doibase 10.1103/PhysRevD.110.114009} {\bibfield  {journal} {\bibinfo
  {journal} {Phys. Rev. D}\ }\textbf {\bibinfo {volume} {110}},\ \bibinfo
  {pages} {114009} (\bibinfo {year} {2024})}\BibitemShut {NoStop}%
\bibitem [{\citenamefont {Dutra}\ \emph {et~al.}(2014)\citenamefont {Dutra},
  \citenamefont {Louren{\c{c}}o}, \citenamefont {Avancini}, \citenamefont
  {Carlson}, \citenamefont {Delfino}, \citenamefont {Menezes}, \citenamefont
  {Provid{\^e}ncia}, \citenamefont {Typel},\ and\ \citenamefont
  {Stone}}]{Dutra2014RMFConstraints}%
  \BibitemOpen
  \bibfield  {author} {\bibinfo {author} {\bibfnamefont {M.}~\bibnamefont
  {Dutra}}, \bibinfo {author} {\bibfnamefont {O.}~\bibnamefont
  {Louren{\c{c}}o}}, \bibinfo {author} {\bibfnamefont {S.~S.}\ \bibnamefont
  {Avancini}}, \bibinfo {author} {\bibfnamefont {B.~V.}\ \bibnamefont
  {Carlson}}, \bibinfo {author} {\bibfnamefont {A.}~\bibnamefont {Delfino}},
  \bibinfo {author} {\bibfnamefont {D.~P.}\ \bibnamefont {Menezes}}, \bibinfo
  {author} {\bibfnamefont {C.}~\bibnamefont {Provid{\^e}ncia}}, \bibinfo
  {author} {\bibfnamefont {S.}~\bibnamefont {Typel}}, \ and\ \bibinfo {author}
  {\bibfnamefont {J.~R.}\ \bibnamefont {Stone}},\ }\href {\doibase
  10.1103/PhysRevC.90.055203} {\bibfield  {journal} {\bibinfo  {journal} {Phys.
  Rev. C}\ }\textbf {\bibinfo {volume} {90}},\ \bibinfo {pages} {055203}
  (\bibinfo {year} {2014})},\ \Eprint {http://arxiv.org/abs/1405.3633}
  {arXiv:1405.3633} \BibitemShut {NoStop}%
\bibitem [{\citenamefont {Li}\ \emph {et~al.}(2008)\citenamefont {Li},
  \citenamefont {Chen},\ and\ \citenamefont {Ko}}]{LiChenKo2008IsospinPR}%
  \BibitemOpen
  \bibfield  {author} {\bibinfo {author} {\bibfnamefont {B.-A.}\ \bibnamefont
  {Li}}, \bibinfo {author} {\bibfnamefont {L.-W.}\ \bibnamefont {Chen}}, \ and\
  \bibinfo {author} {\bibfnamefont {C.~M.}\ \bibnamefont {Ko}},\ }\href
  {\doibase 10.1016/j.physrep.2008.04.005} {\bibfield  {journal} {\bibinfo
  {journal} {Phys. Rep.}\ }\textbf {\bibinfo {volume} {464}},\ \bibinfo {pages}
  {113} (\bibinfo {year} {2008})},\ \Eprint {http://arxiv.org/abs/0804.3580}
  {arXiv:0804.3580} \BibitemShut {NoStop}%
\bibitem [{\citenamefont {Tsang}\ \emph {et~al.}(2012)\citenamefont {Tsang},
  \citenamefont {Stone}, \citenamefont {Camera}, \citenamefont {Danielewicz},
  \citenamefont {Gandolfi}, \citenamefont {Hebeler}, \citenamefont {Horowitz},
  \citenamefont {Lee}, \citenamefont {Lynch}, \citenamefont {Kohley},
  \citenamefont {Lemmon}, \citenamefont {Moller}, \citenamefont {Murakami},
  \citenamefont {Riordan}, \citenamefont {Roca-Maza}, \citenamefont
  {Sammarruca}, \citenamefont {Steiner}, \citenamefont {Vida{\~n}a},\ and\
  \citenamefont {Yennello}}]{Tsang2012SymConstraints}%
  \BibitemOpen
  \bibfield  {author} {\bibinfo {author} {\bibfnamefont {M.~B.}\ \bibnamefont
  {Tsang}}, \bibinfo {author} {\bibfnamefont {J.~R.}\ \bibnamefont {Stone}},
  \bibinfo {author} {\bibfnamefont {F.}~\bibnamefont {Camera}}, \bibinfo
  {author} {\bibfnamefont {P.}~\bibnamefont {Danielewicz}}, \bibinfo {author}
  {\bibfnamefont {S.}~\bibnamefont {Gandolfi}}, \bibinfo {author}
  {\bibfnamefont {K.}~\bibnamefont {Hebeler}}, \bibinfo {author} {\bibfnamefont
  {C.~J.}\ \bibnamefont {Horowitz}}, \bibinfo {author} {\bibfnamefont
  {J.}~\bibnamefont {Lee}}, \bibinfo {author} {\bibfnamefont {W.~G.}\
  \bibnamefont {Lynch}}, \bibinfo {author} {\bibfnamefont {Z.}~\bibnamefont
  {Kohley}}, \bibinfo {author} {\bibfnamefont {R.}~\bibnamefont {Lemmon}},
  \bibinfo {author} {\bibfnamefont {P.}~\bibnamefont {Moller}}, \bibinfo
  {author} {\bibfnamefont {T.}~\bibnamefont {Murakami}}, \bibinfo {author}
  {\bibfnamefont {S.}~\bibnamefont {Riordan}}, \bibinfo {author} {\bibfnamefont
  {X.}~\bibnamefont {Roca-Maza}}, \bibinfo {author} {\bibfnamefont
  {F.}~\bibnamefont {Sammarruca}}, \bibinfo {author} {\bibfnamefont {A.~W.}\
  \bibnamefont {Steiner}}, \bibinfo {author} {\bibfnamefont {I.}~\bibnamefont
  {Vida{\~n}a}}, \ and\ \bibinfo {author} {\bibfnamefont {S.~J.}\ \bibnamefont
  {Yennello}},\ }\href {\doibase 10.1103/PhysRevC.86.015803} {\bibfield
  {journal} {\bibinfo  {journal} {Phys. Rev. C}\ }\textbf {\bibinfo {volume}
  {86}},\ \bibinfo {pages} {015803} (\bibinfo {year} {2012})},\ \Eprint
  {http://arxiv.org/abs/1204.0466} {arXiv:1204.0466} \BibitemShut {NoStop}%
\bibitem [{\citenamefont {Douchin}\ and\ \citenamefont
  {Haensel}(2001)}]{DouchinHaensel2001SLy4}%
  \BibitemOpen
  \bibfield  {author} {\bibinfo {author} {\bibfnamefont {F.}~\bibnamefont
  {Douchin}}\ and\ \bibinfo {author} {\bibfnamefont {P.}~\bibnamefont
  {Haensel}},\ }\href {\doibase 10.1051/0004-6361:20011402} {\bibfield
  {journal} {\bibinfo  {journal} {Astron. Astrophys.}\ }\textbf {\bibinfo
  {volume} {380}},\ \bibinfo {pages} {151} (\bibinfo {year} {2001})},\ \Eprint
  {http://arxiv.org/abs/astro-ph/0111092} {arXiv:astro-ph/0111092} \BibitemShut
  {NoStop}%
\bibitem [{\citenamefont {Tolman}(1939)}]{Tolman1939}%
  \BibitemOpen
  \bibfield  {author} {\bibinfo {author} {\bibfnamefont {R.~C.}\ \bibnamefont
  {Tolman}},\ }\href {\doibase 10.1103/PhysRev.55.364} {\bibfield  {journal}
  {\bibinfo  {journal} {Phys. Rev.}\ }\textbf {\bibinfo {volume} {55}},\
  \bibinfo {pages} {364} (\bibinfo {year} {1939})}\BibitemShut {NoStop}%
\bibitem [{\citenamefont {Xie}\ and\ \citenamefont
  {Li}(2019)}]{Xie2019_ApJ883-174}%
  \BibitemOpen
  \bibfield  {author} {\bibinfo {author} {\bibfnamefont {W.-J.}\ \bibnamefont
  {Xie}}\ and\ \bibinfo {author} {\bibfnamefont {B.-A.}\ \bibnamefont {Li}},\
  }\href {\doibase 10.3847/1538-4357/ab3f37} {\bibfield  {journal} {\bibinfo
  {journal} {Astrophys. J.}\ }\textbf {\bibinfo {volume} {883}},\ \bibinfo
  {pages} {174} (\bibinfo {year} {2019})}\BibitemShut {NoStop}%
\bibitem [{\citenamefont {Xie}\ \emph {et~al.}(2024)\citenamefont {Xie},
  \citenamefont {Li},\ and\ \citenamefont {Zhang}}]{Xie2024PRD110_043025}%
  \BibitemOpen
  \bibfield  {author} {\bibinfo {author} {\bibfnamefont {W.-J.}\ \bibnamefont
  {Xie}}, \bibinfo {author} {\bibfnamefont {B.-A.}\ \bibnamefont {Li}}, \ and\
  \bibinfo {author} {\bibfnamefont {N.-B.}\ \bibnamefont {Zhang}},\ }\href
  {\doibase 10.1103/PhysRevD.110.043025} {\bibfield  {journal} {\bibinfo
  {journal} {Phys. Rev. D}\ }\textbf {\bibinfo {volume} {110}},\ \bibinfo
  {pages} {043025} (\bibinfo {year} {2024})}\BibitemShut {NoStop}%
\bibitem [{\citenamefont {Xie}\ \emph {et~al.}(2026)\citenamefont {Xie},
  \citenamefont {Xia}, \citenamefont {Zhang},\ and\ \citenamefont
  {Xu}}]{xie2026prd}%
  \BibitemOpen
  \bibfield  {author} {\bibinfo {author} {\bibfnamefont {W.-J.}\ \bibnamefont
  {Xie}}, \bibinfo {author} {\bibfnamefont {C.-J.}\ \bibnamefont {Xia}},
  \bibinfo {author} {\bibfnamefont {C.}~\bibnamefont {Zhang}}, \ and\ \bibinfo
  {author} {\bibfnamefont {R.}~\bibnamefont {Xu}},\ }\href
  {https://arxiv.org/abs/2506.22781} {\enquote {\bibinfo {title} {Bayesian
  constraints on quark stars from multi-messenger observations},}\ } (\bibinfo
  {year} {2026}),\ \Eprint {http://arxiv.org/abs/2506.22781} {arXiv:2506.22781
  [astro-ph.HE]} \BibitemShut {NoStop}%
\bibitem [{\citenamefont {Skilling}(2006)}]{Skilling2006NestedSampling}%
  \BibitemOpen
  \bibfield  {author} {\bibinfo {author} {\bibfnamefont {J.}~\bibnamefont
  {Skilling}},\ }\href {\doibase 10.1214/06-BA127} {\bibfield  {journal}
  {\bibinfo  {journal} {Bayesian Analysis}\ }\textbf {\bibinfo {volume} {1}},\
  \bibinfo {pages} {833} (\bibinfo {year} {2006})}\BibitemShut {NoStop}%
\bibitem [{\citenamefont {Antoniadis}\ \emph {et~al.}(2013)\citenamefont
  {Antoniadis}, \citenamefont {Freire}, \citenamefont {Wex}, \citenamefont
  {Tauris}, \citenamefont {Lynch}, \citenamefont {van Kerkwijk}, \citenamefont
  {Kramer}, \citenamefont {Bassa}, \citenamefont {Dhillon}, \citenamefont
  {Driebe}, \citenamefont {Hessels}, \citenamefont {Kaspi}, \citenamefont
  {Kondratiev}, \citenamefont {Langer}, \citenamefont {Marsh}, \citenamefont
  {McLaughlin}, \citenamefont {Pennucci}, \citenamefont {Ransom}, \citenamefont
  {Stairs}, \citenamefont {van Leeuwen}, \citenamefont {Verbiest},\ and\
  \citenamefont {Whelan}}]{Antoniadis2013J0348}%
  \BibitemOpen
  \bibfield  {author} {\bibinfo {author} {\bibfnamefont {J.}~\bibnamefont
  {Antoniadis}}, \bibinfo {author} {\bibfnamefont {P.~C.~C.}\ \bibnamefont
  {Freire}}, \bibinfo {author} {\bibfnamefont {N.}~\bibnamefont {Wex}},
  \bibinfo {author} {\bibfnamefont {T.~M.}\ \bibnamefont {Tauris}}, \bibinfo
  {author} {\bibfnamefont {R.~S.}\ \bibnamefont {Lynch}}, \bibinfo {author}
  {\bibfnamefont {M.~H.}\ \bibnamefont {van Kerkwijk}}, \bibinfo {author}
  {\bibfnamefont {M.}~\bibnamefont {Kramer}}, \bibinfo {author} {\bibfnamefont
  {C.}~\bibnamefont {Bassa}}, \bibinfo {author} {\bibfnamefont {V.~S.}\
  \bibnamefont {Dhillon}}, \bibinfo {author} {\bibfnamefont {T.}~\bibnamefont
  {Driebe}}, \bibinfo {author} {\bibfnamefont {J.~W.~T.}\ \bibnamefont
  {Hessels}}, \bibinfo {author} {\bibfnamefont {V.~M.}\ \bibnamefont {Kaspi}},
  \bibinfo {author} {\bibfnamefont {V.~I.}\ \bibnamefont {Kondratiev}},
  \bibinfo {author} {\bibfnamefont {N.}~\bibnamefont {Langer}}, \bibinfo
  {author} {\bibfnamefont {T.~R.}\ \bibnamefont {Marsh}}, \bibinfo {author}
  {\bibfnamefont {M.~A.}\ \bibnamefont {McLaughlin}}, \bibinfo {author}
  {\bibfnamefont {T.~T.}\ \bibnamefont {Pennucci}}, \bibinfo {author}
  {\bibfnamefont {S.~M.}\ \bibnamefont {Ransom}}, \bibinfo {author}
  {\bibfnamefont {I.~H.}\ \bibnamefont {Stairs}}, \bibinfo {author}
  {\bibfnamefont {J.}~\bibnamefont {van Leeuwen}}, \bibinfo {author}
  {\bibfnamefont {J.~P.~W.}\ \bibnamefont {Verbiest}}, \ and\ \bibinfo {author}
  {\bibfnamefont {D.~G.}\ \bibnamefont {Whelan}},\ }\href {\doibase
  10.1126/science.1233232} {\bibfield  {journal} {\bibinfo  {journal}
  {Science}\ }\textbf {\bibinfo {volume} {340}},\ \bibinfo {pages} {448}
  (\bibinfo {year} {2013})},\ \Eprint {http://arxiv.org/abs/1304.6875}
  {arXiv:1304.6875} \BibitemShut {NoStop}%
\bibitem [{\citenamefont {Danielewicz}\ \emph
  {et~al.}(2002{\natexlab{b}})\citenamefont {Danielewicz}, \citenamefont
  {Lacey},\ and\ \citenamefont {Lynch}}]{Danielewicz2002_Science298-1592}%
  \BibitemOpen
  \bibfield  {author} {\bibinfo {author} {\bibfnamefont {P.}~\bibnamefont
  {Danielewicz}}, \bibinfo {author} {\bibfnamefont {R.}~\bibnamefont {Lacey}},
  \ and\ \bibinfo {author} {\bibfnamefont {W.~G.}\ \bibnamefont {Lynch}},\
  }\href {\doibase 10.1126/science.1078070} {\bibfield  {journal} {\bibinfo
  {journal} {Science}\ }\textbf {\bibinfo {volume} {298}},\ \bibinfo {pages}
  {1592} (\bibinfo {year} {2002}{\natexlab{b}})}\BibitemShut {NoStop}%
\bibitem [{\citenamefont {Kass}\ and\ \citenamefont
  {Raftery}(1995)}]{KassRaftery1995}%
  \BibitemOpen
  \bibfield  {author} {\bibinfo {author} {\bibfnamefont {R.~E.}\ \bibnamefont
  {Kass}}\ and\ \bibinfo {author} {\bibfnamefont {A.~E.}\ \bibnamefont
  {Raftery}},\ }\href {\doibase 10.1080/01621459.1995.10476572} {\bibfield
  {journal} {\bibinfo  {journal} {J. Am. Stat. Assoc.}\ }\textbf {\bibinfo
  {volume} {90}},\ \bibinfo {pages} {773} (\bibinfo {year} {1995})}\BibitemShut
  {NoStop}%
\bibitem [{\citenamefont {Trotta}(2008)}]{Trotta2008BayesSky}%
  \BibitemOpen
  \bibfield  {author} {\bibinfo {author} {\bibfnamefont {R.}~\bibnamefont
  {Trotta}},\ }\href {\doibase 10.1080/00107510802066753} {\bibfield  {journal}
  {\bibinfo  {journal} {Contemporary Physics}\ }\textbf {\bibinfo {volume}
  {49}},\ \bibinfo {pages} {71} (\bibinfo {year} {2008})},\ \Eprint
  {http://arxiv.org/abs/0803.4089} {arXiv:0803.4089} \BibitemShut {NoStop}%
\bibitem [{\citenamefont {Shlomo}\ \emph {et~al.}(2006)\citenamefont {Shlomo},
  \citenamefont {Kolomietz},\ and\ \citenamefont
  {Col{\`o}}}]{Shlomo2006_EPJA30-23}%
  \BibitemOpen
  \bibfield  {author} {\bibinfo {author} {\bibfnamefont {S.}~\bibnamefont
  {Shlomo}}, \bibinfo {author} {\bibfnamefont {V.~M.}\ \bibnamefont
  {Kolomietz}}, \ and\ \bibinfo {author} {\bibfnamefont {G.}~\bibnamefont
  {Col{\`o}}},\ }\href {\doibase 10.1140/epja/i2006-10100-3} {\bibfield
  {journal} {\bibinfo  {journal} {Eur. Phys. J. A}\ }\textbf {\bibinfo {volume}
  {30}},\ \bibinfo {pages} {23} (\bibinfo {year} {2006})}\BibitemShut {NoStop}%
\bibitem [{\citenamefont {Li}\ \emph {et~al.}(2023)\citenamefont {Li},
  \citenamefont {Niu},\ and\ \citenamefont {Col{\`o}}}]{Li2023PRL_ISGMR}%
  \BibitemOpen
  \bibfield  {author} {\bibinfo {author} {\bibfnamefont {Z.~Z.}\ \bibnamefont
  {Li}}, \bibinfo {author} {\bibfnamefont {Y.~F.}\ \bibnamefont {Niu}}, \ and\
  \bibinfo {author} {\bibfnamefont {G.}~\bibnamefont {Col{\`o}}},\ }\href
  {\doibase 10.1103/PhysRevLett.131.082501} {\bibfield  {journal} {\bibinfo
  {journal} {Phys. Rev. Lett.}\ }\textbf {\bibinfo {volume} {131}},\ \bibinfo
  {pages} {082501} (\bibinfo {year} {2023})},\ \Eprint
  {http://arxiv.org/abs/2211.01264} {arXiv:2211.01264 [nucl-th]} \BibitemShut
  {NoStop}%
\bibitem [{\citenamefont {Zhou}\ \emph {et~al.}(2023)\citenamefont {Zhou},
  \citenamefont {Xu},\ and\ \citenamefont
  {Papakonstantinou}}]{Zhou2023PRC_NSInference}%
  \BibitemOpen
  \bibfield  {author} {\bibinfo {author} {\bibfnamefont {J.}~\bibnamefont
  {Zhou}}, \bibinfo {author} {\bibfnamefont {J.}~\bibnamefont {Xu}}, \ and\
  \bibinfo {author} {\bibfnamefont {P.}~\bibnamefont {Papakonstantinou}},\
  }\href {\doibase 10.1103/PhysRevC.107.055803} {\bibfield  {journal} {\bibinfo
   {journal} {Phys. Rev. C}\ }\textbf {\bibinfo {volume} {107}},\ \bibinfo
  {pages} {055803} (\bibinfo {year} {2023})},\ \Eprint
  {http://arxiv.org/abs/2301.07904} {arXiv:2301.07904 [nucl-th]} \BibitemShut
  {NoStop}%
\bibitem [{\citenamefont {Xu}\ \emph {et~al.}(2021)\citenamefont {Xu},
  \citenamefont {Zhang},\ and\ \citenamefont {Li}}]{Xu2021PRC_K0Bayes}%
  \BibitemOpen
  \bibfield  {author} {\bibinfo {author} {\bibfnamefont {J.}~\bibnamefont
  {Xu}}, \bibinfo {author} {\bibfnamefont {Z.}~\bibnamefont {Zhang}}, \ and\
  \bibinfo {author} {\bibfnamefont {B.-A.}\ \bibnamefont {Li}},\ }\href
  {\doibase 10.1103/PhysRevC.104.054324} {\bibfield  {journal} {\bibinfo
  {journal} {Phys. Rev. C}\ }\textbf {\bibinfo {volume} {104}},\ \bibinfo
  {pages} {054324} (\bibinfo {year} {2021})},\ \Eprint
  {http://arxiv.org/abs/2107.10962} {arXiv:2107.10962 [nucl-th]} \BibitemShut
  {NoStop}%
\bibitem [{\citenamefont {Oertel}\ \emph
  {et~al.}(2017{\natexlab{b}})\citenamefont {Oertel}, \citenamefont {Hempel},
  \citenamefont {Kl\"ahn},\ and\ \citenamefont
  {Typel}}]{Oertel2017_RMP89-015007}%
  \BibitemOpen
  \bibfield  {author} {\bibinfo {author} {\bibfnamefont {M.}~\bibnamefont
  {Oertel}}, \bibinfo {author} {\bibfnamefont {M.}~\bibnamefont {Hempel}},
  \bibinfo {author} {\bibfnamefont {T.}~\bibnamefont {Kl\"ahn}}, \ and\
  \bibinfo {author} {\bibfnamefont {S.}~\bibnamefont {Typel}},\ }\href
  {\doibase 10.1103/RevModPhys.89.015007} {\bibfield  {journal} {\bibinfo
  {journal} {Rev. Mod. Phys.}\ }\textbf {\bibinfo {volume} {89}},\ \bibinfo
  {pages} {015007} (\bibinfo {year} {2017}{\natexlab{b}})}\BibitemShut
  {NoStop}%
\bibitem [{\citenamefont {Margueron}\ \emph {et~al.}(2018)\citenamefont
  {Margueron}, \citenamefont {Hoffmann~Casali},\ and\ \citenamefont
  {Gulminelli}}]{Margueron2018PRC_MetaEOS_NS}%
  \BibitemOpen
  \bibfield  {author} {\bibinfo {author} {\bibfnamefont {J.}~\bibnamefont
  {Margueron}}, \bibinfo {author} {\bibfnamefont {R.}~\bibnamefont
  {Hoffmann~Casali}}, \ and\ \bibinfo {author} {\bibfnamefont {F.}~\bibnamefont
  {Gulminelli}},\ }\href {\doibase 10.1103/PhysRevC.97.025806} {\bibfield
  {journal} {\bibinfo  {journal} {Phys. Rev. C}\ }\textbf {\bibinfo {volume}
  {97}},\ \bibinfo {pages} {025806} (\bibinfo {year} {2018})},\ \Eprint
  {http://arxiv.org/abs/1708.06895} {arXiv:1708.06895 [nucl-th]} \BibitemShut
  {NoStop}%
\bibitem [{\citenamefont {Typel}\ \emph {et~al.}(2010)\citenamefont {Typel},
  \citenamefont {R\"opke}, \citenamefont {Kl\"ahn}, \citenamefont {Blaschke},\
  and\ \citenamefont {Wolter}}]{Typel2010_PRC81-015803}%
  \BibitemOpen
  \bibfield  {author} {\bibinfo {author} {\bibfnamefont {S.}~\bibnamefont
  {Typel}}, \bibinfo {author} {\bibfnamefont {G.}~\bibnamefont {R\"opke}},
  \bibinfo {author} {\bibfnamefont {T.}~\bibnamefont {Kl\"ahn}}, \bibinfo
  {author} {\bibfnamefont {D.}~\bibnamefont {Blaschke}}, \ and\ \bibinfo
  {author} {\bibfnamefont {H.~H.}\ \bibnamefont {Wolter}},\ }\href {\doibase
  10.1103/PhysRevC.81.015803} {\bibfield  {journal} {\bibinfo  {journal} {Phys.
  Rev. C}\ }\textbf {\bibinfo {volume} {81}},\ \bibinfo {pages} {015803}
  (\bibinfo {year} {2010})}\BibitemShut {NoStop}%
\bibitem [{\citenamefont {Annala}\ \emph
  {et~al.}(2020{\natexlab{b}})\citenamefont {Annala}, \citenamefont {Gorda},
  \citenamefont {Kurkela}, \citenamefont {N\"attil\"a},\ and\ \citenamefont
  {Vuorinen}}]{Annala2020_NP}%
  \BibitemOpen
  \bibfield  {author} {\bibinfo {author} {\bibfnamefont {E.}~\bibnamefont
  {Annala}}, \bibinfo {author} {\bibfnamefont {T.}~\bibnamefont {Gorda}},
  \bibinfo {author} {\bibfnamefont {A.}~\bibnamefont {Kurkela}}, \bibinfo
  {author} {\bibfnamefont {J.}~\bibnamefont {N\"attil\"a}}, \ and\ \bibinfo
  {author} {\bibfnamefont {A.}~\bibnamefont {Vuorinen}},\ }\href {\doibase
  10.1038/s41567-020-0914-9} {\bibfield  {journal} {\bibinfo  {journal} {Nat.
  Phys.}\ }\textbf {\bibinfo {volume} {16}},\ \bibinfo {pages} {907} (\bibinfo
  {year} {2020}{\natexlab{b}})}\BibitemShut {NoStop}%
\bibitem [{\citenamefont {Tews}\ \emph {et~al.}(2018)\citenamefont {Tews},
  \citenamefont {Carlson}, \citenamefont {Gandolfi},\ and\ \citenamefont
  {Reddy}}]{Tews2018ApJ}%
  \BibitemOpen
  \bibfield  {author} {\bibinfo {author} {\bibfnamefont {I.}~\bibnamefont
  {Tews}}, \bibinfo {author} {\bibfnamefont {J.}~\bibnamefont {Carlson}},
  \bibinfo {author} {\bibfnamefont {S.}~\bibnamefont {Gandolfi}}, \ and\
  \bibinfo {author} {\bibfnamefont {S.}~\bibnamefont {Reddy}},\ }\href
  {\doibase 10.3847/1538-4357/aac267} {\bibfield  {journal} {\bibinfo
  {journal} {The Astrophysical Journal}\ }\textbf {\bibinfo {volume} {860}},\
  \bibinfo {pages} {149} (\bibinfo {year} {2018})}\BibitemShut {NoStop}%
\bibitem [{\citenamefont {Reed}\ and\ \citenamefont
  {Horowitz}(2020)}]{Reed2020PRC}%
  \BibitemOpen
  \bibfield  {author} {\bibinfo {author} {\bibfnamefont {B.}~\bibnamefont
  {Reed}}\ and\ \bibinfo {author} {\bibfnamefont {C.~J.}\ \bibnamefont
  {Horowitz}},\ }\href {\doibase 10.1103/PhysRevC.101.045803} {\bibfield
  {journal} {\bibinfo  {journal} {Physical Review C}\ }\textbf {\bibinfo
  {volume} {101}},\ \bibinfo {pages} {045803} (\bibinfo {year}
  {2020})}\BibitemShut {NoStop}%
\end{thebibliography}
\end{document}